\documentclass[twocolumn]{aastex631}

\begin{document}

\title{Shock breakouts from compact CSM surrounding core-collapse SN progenitors may contribute significantly to the observed $\gtrsim10$~TeV neutrino background}

\shorttitle{SN breakouts and the neutrino background}
\shortauthors{Waxman et al.}

\author[0000-0002-9038-5877]{Eli Waxman}
\affiliation{Dept. of Particle Phys. \& Astrophys., Weizmann Institute of Science, Rehovot 76100, Israel}

\author{Tal Wasserman}
\affiliation{Dept. of Particle Phys. \& Astrophys., Weizmann Institute of Science, Rehovot 76100, Israel}

\author{Eran O. Ofek}
\affiliation{Dept. of Particle Phys. \& Astrophys., Weizmann Institute of Science, Rehovot 76100, Israel}

\author{Avishay Gal-Yam}
\affiliation{Dept. of Particle Phys. \& Astrophys., Weizmann Institute of Science, Rehovot 76100, Israel}

\begin{abstract}
Growing observational evidence suggests that enhanced mass loss from the progenitors of core-collapse supernovae (SNe) is common during $\sim1$~yr preceding the explosion, creating an optically thick circum-stellar medium (CSM) shell at $\sim10^{14.5}$~cm radii. We show that if such mass loss is indeed common, then the breakout of the SN shock through the dense CSM shell produces a neutrino flux that may account for a significant fraction of the observed $\gtrsim10$~TeV neutrino background. The neutrinos are created within a few days from the explosion, during and shortly after the shock breakout, which produces also large UV (and later X-ray) luminosity. The compact size and large UV luminosity imply a pair production optical depth of $\sim10^4$ for $>100$~GeV photons, naturally accounting for the lack of a high-energy gamma-ray background accompanying the neutrino background. SNe producing $>1$ neutrino event in a 1~km$^2$ detector are expected at a rate of $\lesssim0.1$/yr. A quantitative theory describing the evolution of the electromagnetic spectrum during a breakout, as the radiation-mediated shock is transformed into a collisionless one, is required to enable (i) using data from upcoming surveys that will systematically detect large numbers of young, $<1$~d old SNe, to determine the pre-explosion mass loss history of the SN progenitor population, and (ii) a quantitative determination of the neutrino luminosity and spectrum. 
\end{abstract}

\section{Introduction}
\label{sec:Intro}

The origin of the astrophysical high energy, $\gtrsim10$~TeV, neutrino background \citep{2013IC-detection,2019ANTARES-diffuse} is unknown. The isotropy of the background implies that extra-galactic sources dominate it. The fact that the intensity is similar to the Waxman-Bahcall (WB) bound \citep{1998WB} means that the neutrino energy production rate is similar to the rate of production of ultra-high energy, $>10^{19}$~eV cosmic-rays (CRs), suggesting that the neutrino and ultra-high energy CR sources may be related. However, we do not have direct experimental evidence linking the neutrino and cosmic-ray sources, which remain unknown \citep[see][for a recent comprehensive review]{2022Snowmass-nu}. 

The reported 10--30\,TeV neutrino intensity, $\varepsilon_\nu^2 dI/d\varepsilon_\nu\approx10^{-7}$\,GeV\,cm$^{-2}$\,s$^{-1}$\,sr$^{-1}$ is approximately twice the WB bound and is similar to the $\sim0.1$\,TeV gamma-ray background \citep{2015Fermi-XGB}. The latter coincidence implies some tension between neutrino and gamma-ray observations \citep{2015Ando-g-constraint,2016Murase-g-tension}. This is because high energy neutrino production by pion decay is accompanied by high energy gamma-ray production at a similar rate, leading (after electromagnetic cascades due to interaction with the IR background) to a $\sim0.1$\,TeV gamma-ray intensity similar to the observed background, which is already largely accounted for by emission from blazars \citep{2016Fermi-gbgnd-blazars}, that were found not to dominate the neutrino background \citep{2017IC-blazar-limit}. This tension led to increased interest in sources that are opaque to high-energy, $>0.1$~TeV gamma-rays, particularly "choked-jets" in exploding massive stars \citep{2001MW-chocked,2013Murase-chocked,2018Tamborra-chocked,2024Zegarelli-choked-jets}, hyper-Eddington accretion onto compact objects \citep{2024Metzger-FRB-nu}, and the hot (X-ray) coronae expected to exist near 
%(at $\sim10$ Schwartzschild radii distance from) 
the accreting super-massive black holes of active galactic nuclei (AGN) \citep{1991Stecker-AGN-core,2015Semikoz-AGN-core,2020Murase-AGN-core}. The latter hypothesis is supported by the 4$\sigma$ association \citep{2022IC-1068} of 1--10\,TeV neutrinos with the NGC1068 Seyfert galaxy. However, the models for neutrino emission from the X-ray corona are challenged by the fact that a very large CR luminosity, amounting to $\approx0.1$ the Eddington luminosity, is required to be produced within a narrow CR energy range, 10--30~TeV, and by the fact that the acceleration mechanism is uncertain \citep{2024Murase-1068,2024Resconi-1068}.

Here, we discuss an alternative origin for the $\gtrsim10$~TeV neutrino background -- SN shocks breaking through compact optically thick CSM surrounding SN progenitors. Collisionless SN shocks driven into the winds surrounding massive stars and into the interstellar medium (ISM) have long been suggested as sources of CRs \citep{1987BlandfordEichler}, and the high energy neutrino emission due to energy losses by these CRs was discussed by several authors \citep[e.g.][]{2011Murase-SNe,2017Petropoulou-IIn-Nu,2022Tamborra-SNe}. The neutrino luminosity produced by shocks driven into typical wind/ISM extends over hundreds of days. It is too low to account for the neutrino background, and it is accompanied by high energy gamma-ray luminosity similar to the neutrino luminosity since the pair production optical depth is low for most of the emission, that occurs at large, $>10^{16}$~cm radii \citep[see the end of \S~\ref{sec:BO} and \S~\ref{sec:discussion} for a short discussion of the recent work of][]{2022Tamborra-SNe,2024Murase-SNe}. %, and is undetectable for individual SNe by 1~km$^2$ effective area neutrino detectors. 

We have shown \citep[][hereafter KSW11]{2011KSW11} that if a sufficiently optically thick CSM surrounds the SN progenitor star, then a large neutrino luminosity may be produced during and shortly after the SN shock breakout through the dense CSM shell, i.e. during and shortly after the shock transition from being radiation-mediated to collisionless \citep[see also][]{2019Zhuo-PeV-BO}. Our motivation for considering the presence of such dense CSM was the observation of several X-ray flashes and low-luminosity gamma-ray bursts associated with SNe, which may be explained as (fast) shock breakouts from optically thick compact CSM. We return to this topic here since, over the past few years, the observational constraints on stellar mass loss preceding SN explosions have greatly improved, providing for the first time initial information on the prevalence and properties of mass loss in the type II SN progenitor population. While pre-explosion mass loss may also be expected in other rare types of core-collapse SN progenitors, we focus here on the population of type II progenitors, which are hydrogen-rich super-giants, for which observations allow one to draw initial constraints on the population as a whole. The possible contribution of other SN types is briefly discussed in \S~\ref{sec:discussion}. 

The Paper is organized as follows. The observational constraints on compact optically thick CSM surrounding type II SN progenitors are discussed in \S~\ref{sec:mass-loss}. The neutrino energy production rate in the universe, required to account for the observed neutrino background, is derived in \S~\ref{sec:nu-rat}, and the production of neutrinos and radiation in shock breakouts are discussed in \S~\ref{sec:BO}. Our conclusions are summarized, and their implications for future experimental and theoretical work are discussed in \S~\ref{sec:discussion}.

\section{Pre-explosion mass loss, neutrinos, and shock breakout}
\label{sec:analysis}

\subsection{Pre-explosion mass loss of type II SN progenitors}
\label{sec:mass-loss}

Systematic analyses \citep{Ofek+2014_IIn_precursors,2021Nora-IIn-precursors} of precursor emission (preceding the SN explosion) in a large sample of SNe of Type IIn, i.e., showing narrow line spectra indicative of CSM interaction \citep{Schlegel1990, Smith2014ARA&A_TypeIIn_SN_MassLoss,Gal-Yam2017}, find that significant precursor emission, with optical photon energy exceeding $10^{47}$\,erg, is common during the 90 days preceding the SN explosion. The mechanism producing the precursors is not well understood \citep[see some suggestions in][]{2007Woosley-PI-Mloss,2012Chevalier-binary-Mloss,Soker13CommonE,Suarez13Mdot,2014Smith-Mloss,Shiode+Quataert2014_WaveDrivenMassLoss,Fuller2017_precursors_RSG, Fuller+2018_precursors}. However, the precursor probably reflects the deposition of energy (exceeding the observed optical photon energy) at the stellar envelope. Since $10^{47}$\,erg corresponds to the binding energy of 1 solar mass at the envelope of a red-super-giant (RSG), $GM_*M/R_*$ with $M_*=15M_\odot,\,M=1\,M_\odot,R_*=10^{13.5}\,{\rm cm}$, and since the energy deposition duration is shorter than or comparable to the dynamical time of the envelope, $1/\sqrt{G\rho}\approx100$\,d, the precursor is expected to be associated with the ejection of a significant fraction of a solar mass.
The ejected mass is expected to expand at a velocity comparable to the escape velocity, $\sqrt{2GM_*/R_*}\approx100\,{\rm km{\rm\, s^{-1}}}$, implying that a shell ejected at time $t_{\rm pr}$ preceding the explosion expands to a radius of $\approx10^{14.5}(t_{\rm pr}/1{\rm yr})$~cm by the explosion time. Indeed, \citet{2021Nora-IIn-precursors}  infer a dense 0.1--1\,$M_\odot$ CSM out to $10^{14}-10^{15}$\,cm radii for most of their sample. Interestingly, \citet{Jacobson-Galan+2022ApJ_SN2020tlf_precursor_MassLoss} report the detection of a precursor prior to a spectroscopically-regular Type II SN.

Independent evidence for the prevalent presence of optically thick CSM shells around spectroscopically regular type II SN progenitors is obtained from early, 1\,day time scale observations of optical-UV SN light curves. \citet{2024Irani-Early-UV-SNII} carried out the first systematic analysis of early ($\sim1$\,d) optical-UV light curves of a large sample of type II SNe \citep[see][for a discussion of analyses of individual SNe]{2023Morag-RSG-I}. They find that while the early light curves of $\approx50\%$ of type II SN are consistent with the emission from the expanding shocked stellar envelope, so-called ``shock cooling" emission \citep[see][for reviews]{2017WK-SN-book,2020Levinson-Nakar-rev}, the light curves of the other 50\% are inconsistent with shock cooling and indicate the presence of an optically thick CSM shell. The extended, days long, rise of the luminosity to high values of $10^{43}-10^{44}$\,erg\,s$^{-1}$ and the high color temperature, with most emission in the UV, are consistent with shock breakout from such a shell \citep[][and \S~\ref{sec:BO}]{2010Ofek-wind-BO,2017WK-SN-book}: The shock breakout radius is inferred from the duration of the luminosity rise, which is given by the supernova driven shock crossing time, $R_{\rm br}=vt=10^{14.5}(t/3\,{\rm d})(v/10^4{\rm km{\rm\, s^{-1}}})$ where $v$ is the shock velocity; The shell mass within $R_{\rm br}$ is determined by the requirement that the optical depth at $R_{\rm br}$ equals $c/v$, $M=(c/v)4\pi R_{\rm br}^2/\kappa\approx0.05M_\odot$; The observed breakout energy, $\approx10^{49}$\,erg is consistent with $0.5Mv^2=5\times10^{49}$\,erg; The high, $\sim10$\,eV temperature is consistent with a blackbody radiation carrying the breakout energy at the breakout radius. This work supports earlier suggestions from analyses of the rise time of Type II SN light curves \citep[e.g.][]{Forster2018,2018Morozova-SNII-Mloss-lightcurve} that many of these explosions occur within a compact distribution of CSM.

Additional independent evidence for the existence of an optically thick $\sim10^{14.5}$\,cm CSM shell is provided by ``flash spectroscopy" \citep{Gal-Yam+2014_SN2013cu_FlashSpectroscopy} which revealed the presence of strong narrow spectral lines from high ionization species that disappear a few days following the SN explosion \citep{Gal-Yam+2014_SN2013cu_FlashSpectroscopy,Khazov+2016_FlashSearchSample, Yaron+2017_PTF13dqy_HighIonization_FlashSPectroscopy, Zhang2020,Bruch+2021_SN_Progenitors_ElevatedMassLoss_FlashSpectroscopy,Terreran2022,Jacobson-Galan2024}. These lines are most naturally explained as due to the ionization and excitation by the breakout UV emission of a compact CSM shell that is swept up by the SN shock on a few days time scale (hence extending to $\sim10^{14.5}$\,cm) with optical depth corresponding to mass loss rates of  $\dot{M}/(v_{\rm w}/100{\rm km{\rm\, s^{-1}}})=10^{-3}-10^{-2}M_\odot/$yr \citep{Yaron+2017_PTF13dqy_HighIonization_FlashSPectroscopy,2017Dessart-flash-wind,2019Boian-flash-wind}. \cite{Bruch2023} find that $>50\%$ of SNe II likely show such CSM features.

Finally, recent observations of the nearby 6.4\,Mpc distance SN\,2003ixf enabled an unprecedentedly detailed study of the CSM structure around a super-giant SN progenitor. Early multi-wavelength (optical-UV-X-ray) and spectra measurements were successfully carried out thanks to the early detection and relatively short distance, providing stringent constraints on the CSM at the progenitor's vicinity \citep[e.g.,][]{2023Bostroem-ixf,2023Jacobson-ixf,2023Grefenstette-ixf}. The observations are consistent with a shock breakout through a dense CSM shell, with $\dot{M}/(v_{\rm w}/100{\rm km{\rm\, s^{-1}}})=0.03M_\odot/$yr extending to $\approx2\times10^{14}$~cm, surrounded by a much lower density wind, $\dot{M}/(v_{\rm w}/100{\rm km{\rm\, s^{-1}}})\approx10^{-4}M_\odot/$yr, at larger radii \citep{2024Zimmerman-ixf}. 

Existing analyses do not enable the determination of the density structure within the CSM shell. Observations are consistent with a $1/r^2$ density dependence, which is expected, e.g., for a steady wind mass loss. The results of numerical simulations of mass ejection following energy deposition in the envelope of a giant star yield density distributions, which are not very different from $1/r^2$ \citep[e.g.][]{2020Kuriyama-CSM-profile,2021Tsuna-CSM-profile,2022Kasen-CSM-profile}. Since our results are not very sensitive to the density profile, we will adopt a $1/r^2$ profile for the following calculations. Future observations and improved breakout theory will enable one to better constrain the density profiles.

For a given CSM shell mass $M_{\rm CSM}$ within $10^{14.5}$\,cm, the corresponding steady wind mass loss is given by $\dot{M}/(v_{\rm w}/100\,{\rm km{\rm\, s^{-1}}})=M_{\rm CSM} {\rm yr}^{-1}$, where $v_{\rm w}$ is the wind velocity. The mass loss rates corresponding to $M_{\rm CSM}=0.01-0.1\,M_\odot$, $\dot{M}/(v_{\rm w}/100{\rm km{\rm\,s^{-1}}})=0.01-0.1M_\odot$\,yr$^{-1}$ are orders of magnitude higher than the mass loss rates typically observed for RSGs, $\dot{M}/(v_{\rm w}/100\,{\rm km{\rm\, s^{-1}}})\le10^{-4}M_\odot$\,yr$^{-1}$ \citep{1988Jager-Mdot,2004Marshall-Mdot,2005Loon-Mdot}. The enhanced mass loss over a year time scale preceding the SN explosion is the reason for the orders of magnitude enhanced neutrino luminosity, compared to that expected for interaction with regular RSG winds.

\subsection{The rate of high energy neutrino production in the local universe}
\label{sec:nu-rat}

High-energy astrophysical neutrinos are most likely produced by the decay of pions that, in turn, are produced in interactions of high-energy cosmic rays with photons and nucleons. The (all flavor) neutrino background intensity that will be produced by complete energy loss of cosmic-ray protons to pion production in $p\gamma$ interactions dominated by the $\Delta$ resonance (where charged and neutral pions are produced at a ratio of 1:1) is given by \citep{1998WB,2013WB}
\begin{equation}
\varepsilon_\nu^2 \frac{dI_{\rm WB}}{d\varepsilon_\nu}=3.4\times10^{-8}\frac{\xi_z}{3}\frac{(\varepsilon_p^2d\dot{n}_p/d\varepsilon_p)_{z=0}}{0.5\times10^{44}{\rm erg/Mpc^3yr}}\frac{\rm GeV}{\rm cm^2 s\, sr},
\label{eq:WB}
\end{equation}
where $\varepsilon_p^2d\dot{n}_p/d\varepsilon_p$ is the cosmic-ray proton energy production rate, $0.5\times10^{44}{\rm erg/Mpc^3yr}$ is the local ($z=0$) energy production rate of ultra-high energy CRs, and  $\xi_z$ is (a dimensionless parameter) of order unity, which depends on the redshift evolution of $\varepsilon_p^2d\dot{n}_p/d\varepsilon_p$. The characteristic neutrino energy is approximately 5\% of the parent proton energy. $\xi_z=3$ is obtained for redshift evolution following that of the star-formation rate or AGN luminosity density, $\propto(1+z)^3$ up to $z=2$ and constant at higher $z$ ($\xi_z=0.6$ for no evolution). For $p\gamma$ interactions at energies higher than that of the $\Delta$ resonance, or $pp(n)$ interactions, the charged to neutral pion ratio may be closer to 2:1, increasing the neutrino intensity by $\approx30\%$.

Using Equation~(\ref{eq:WB}), a local ($z=0$) cosmic-ray proton production rate of 
\begin{equation}
    \label{eq:Qp-diff}
    (\varepsilon_p^2d\dot{n}_p/d\varepsilon_p)_{z=0}=
    10^{44}\frac{\varepsilon_\nu^2 dI/d\varepsilon_\nu}{10^{-7}{\rm GeV/(cm^2s\,sr)}}\frac{\rm erg}{\rm Mpc^3yr}
\end{equation}
is required (at $>100$~TeV) to produce the observed neutrino intensity at $\sim10$~TeV through complete energy loss of protons to pions in $pp(n)$ interactions. The corresponding neutrino production rate is
\begin{equation}
    \label{eq:Qnu-diff}
    (\varepsilon_\nu^2d\dot{n}_\nu/d\varepsilon_\nu)_{z=0}=
    5\times10^{43}\frac{\varepsilon_\nu^2 dI/d\varepsilon_\nu}{10^{-7}{\rm GeV/(cm^2s\,sr)}}\frac{\rm erg}{\rm Mpc^3yr}.
\end{equation}

\subsection{Breakout neutrinos and photons}
\label{sec:BO}

We consider the breakout of the SN shock through a CSM shell with density
\begin{equation}
    \label{eq:rho-r}
    \rho(r)=\frac{\dot{M}}{4\pi r^2 v_{\rm w}}.
\end{equation}
The shocked expanding stellar envelope acts as a "piston" driving a shock through the CSM. For $M_{\rm CSM}\ll1M_\odot$, the shock is driven by the fast, outer $\le1M_\odot$ shell of the envelope, which is typically accelerated to $\ge10^4{\rm km/s}$. The velocity profile of the shock-accelerated outer part of the envelope is well understood theoretically for stellar envelopes in hydro-static equilibrium, as explained below. It should, however, be noted that the envelopes of stars undergoing significant mass ejection shortly before explosion may not be in hydrostatic equilibrium. We do not, therefore, rely in our estimates on a detailed model of the mass-velocity profile and provide the results as a function of the piston velocity. Future observations of CSM shock breakouts will enable us to better constrain the outer mass-velocity profiles.

Numeric calculations and analytic calculations
%, based on an interpolation between the Sedov-von Neumann-Taylor and Sakurai self-similar solutions, 
of shock propagation through an exploding RSG with polytropic hydro-static envelop show \citep{1999MnM-SN-shock} that the final (post-shock expansion) velocity attained by the outer shells containing a fraction $\delta_m<0.1$ of the envelope mass is $v>3\sqrt{E_{\rm ej}/M_{\rm ej}}$ corresponding to $v>10^9\,{\rm cm\,s}^{-1}$ for typical values of $\sqrt{E_{\rm ej}/M_{\rm ej}}\approx3,000\,{\rm km\,s^{-1}}$ where $E_{\rm ej}$ and $M_{\rm ej}$ are the ejecta energy and mass respectively. 
While the outer parts are accelerated to still larger (final) velocity, the velocity dependence on mass is very shallow, $v\propto\delta_m^{-0.11}$. This shallow dependence also implies that the density of the expanding envelope shell is much larger than the density of the CSM - the envelope velocity spread implies a thickness $\Delta r=0.11r$ at expansion radius $r$, while the CSM thickness is $\Delta r=r$. Due to the large density difference, the reverse shock driven back into the envelope is weak, and the shocked CSM plasma is accelerated to the ``piston" velocity of $v\gtrsim10^9\,{\rm cm\,s}^{-1}$.

For the high-density CSM considered, the shock expanding through the shell is initially radiation-mediated. Photons start escaping on a dynamical time scale, and the radiation-mediated shock is dissolved once the Thomson optical depth ahead of the shock is comparable to the optical depth of the shock width, $\tau=c/v_s$ (\citealt{Weaver1976_StructureOfSupernovaeShocks}), at a radius
\begin{equation}
    \label{eq:Rbr}
    R_{\rm br}=\frac{v_s}{c}\frac{\kappa\dot{M}}{4\pi v_{\rm w}}=6\times10^{13}\frac{\dot{M}_{-2}v_{s,9}}{v_{\rm w,7}}\,{\rm cm},
\end{equation}
where $\dot{M}=10^{-2}\dot{M}_{-2}M_\odot$\,yr$^{-1}$, $v_{s}=10^9v_{s,9}{\rm cm/s}$ and $v_{\rm w}=10^7v_{\rm w,7}\,{\rm cm\,s}^{-1}$ (note that the shock velocity is somewhat larger than, $7/6$ times, the ``piston" velocity $v$). The breakout luminosity is 
\begin{equation}
    \label{eq:Lbr}
    L_{\rm br}=\frac{1}{2}\frac{\dot{M}}{v_{\rm w}}v^3=
    3\times10^{43}\frac{\dot{M}_{-2}}{v_{\rm w,7}}v_9^3\,{\rm erg\,s}^{-1},
\end{equation}
(noting that the post-shock thermal and kinetic energies are similar and that the shock is highly radiative, see KSW11),
and the breakout duration is
\begin{equation}
    \label{eq:tbr}
    t_{\rm br}=\frac{R_{\rm br}}{v}=6\times10^4 \frac{\dot{M}_{-2}}{v_{\rm w,7}}\,{\rm s}.
\end{equation}
The post-shock radiation is nearly in thermal equilibrium before transitioning to a collisionless shock, implying a spectrum peaking at UV photon energies. The luminosity following the breakout remains roughly constant since the shocked mass, and hence, energy grows linearly with $r$. For mass loss over a year time scale, we expect the high luminosity to extend to $\sim1\,{\rm yr}(v_{\rm w}/v)\approx 3$\,d.
The radiation spectrum shifts with time to higher energies, reaching the X-ray band at $r=$~a few~$R_{\rm br}$ (Wasserman et al. 2024), as the collisionless shock heats the protons to a temperature of $0.2 v_9^2$\,MeV, and the electrons to a temperature exceeding 60\,keV (KSW11).

The conversion of the radiation-mediated shock to a collisionless one occurs, as shown in KSW11 and supported by numeric calculations (Wasserman et al. 2024), at $r=R_{\rm br}/2$. The collisionless shock is expected to accelerate protons to high energy, exceeding 100~TeV. The acceleration to higher energies may be suppressed due to energy loss by $p\gamma$ interactions with X-ray photons; see eqs. 13 \& 14 of KSW11. The high energy protons lose energy by $pp$ collisions efficiently, on a time scale shorter than the dynamical time $r/v$, up to a radius of $r_{pp}\approx 10 v_9^{-2}R_{\rm br}$ (KSW11). The CSM mass contained within this radius is
\begin{equation}
    M_{pp}=\dot{M}\frac{r_{pp}}{v_{\rm w}}= 0.02 v_9^{-1} \left(\frac{\dot{M}_{-2}}{v_{\rm w,7}}\right)^2 M_\odot.
\end{equation}

Assuming that a fraction $\varepsilon_{\rm CR}$ of the thermal energy is carried by non-thermal relativistic protons, the proton energy lost to neutrino production through pion decay is $E_{\nu}\approx 0.5E_{\pi}\approx \frac{1}{4}\varepsilon_{\rm CR}M_{pp}v^2$. For an accelerated proton spectrum of $\varepsilon_p^2dn_p/d\varepsilon_p=\varepsilon_p^0$ over 6 energy decades, 1~GeV to 1000~TeV, the resulting neutrino production is  
\begin{equation}
    \label{eq:Enu-log}
    \varepsilon_\nu^2\frac{d{n}_\nu}{d\varepsilon_\nu}\approx 
    7\times10^{46}\varepsilon_{\rm CR,-1}v_9\left(\frac{\dot{M}_{-2}}{v_{\rm w,7}}\right)^2\,{\rm erg},
\end{equation}
and the energy carried by $>1$~TeV neutrinos is $E_\nu(>1\,{\rm TeV})\approx7\varepsilon_\nu^2dn_\nu/d\varepsilon_\nu$.
These results are not very sensitive to the exact proton and neutrino spectrum. For $\varepsilon_p^2dn_p/d\varepsilon_p=\varepsilon_p^{-0.2}$, $E_\nu(>1\,{\rm TeV})$ is smaller by a factor of 2 compared to the value derived above. Since $M_{pp}$ cannot exceed the ejected CSM mass, the neutrino energy is limited to
\begin{equation}
    \label{eq:Enu-lim}
    \varepsilon_\nu^2\frac{d{n}_\nu}{d\varepsilon_\nu}\le 
    4\varepsilon_{\rm CR,-1}\frac{M_{\rm CSM}}{0.1M_\odot}v_9^2\times10^{47}{\rm erg}.
\end{equation}
Equality is obtained if the dense CSM mass is contained within the radius $r_{pp}$ of efficient pion production, which will be the case for CSM ejection over $<r_{pp}/v_{\rm w}=2(\dot{M}_{-2}/v_{\rm w,7})v_9^{-1}$~yr preceding the explosion. Note that for large CSM mass, $M_{\rm CSM}\gg0.1M_\odot$, the deceleration of the ejecta that drives the shock is expect to be significant, reducing the velocity below $v_9=1$.

The production of high-energy photons accompanies the production of high-energy neutrinos. At $r=10 R_{\rm br}$, the largest radius at which proton energy loss to pions is efficient, the pair production optical depth for high-energy photons that interact with lower energy photons that carry a significant fraction of the photon energy density ($\approx \rho v^2$), is (KSW11)
\begin{equation}
    \label{eq:tau-gg}
    \tau_{\gamma\gamma}\approx 6 v_9 \frac{h\nu}{m_ec^2}.
\end{equation}
For $h\nu=100$\,GeV photons, interacting with $>10$\,eV (UV) photons that carry most of the energy, $\tau_{\gamma\gamma}>10^5$. For 1\,GeV photons interacting with X-ray photons, an optical depth $>10^3$ would be obtained at $r>$~a few~$R_{\rm br}$, where the emission shifts to the X-ray band.

We note that \citet{2022Tamborra-SNe} predict a neutrino luminosity $\varepsilon_\nu^2d\dot{n}_\nu/d\varepsilon_\nu\approx2\times 10^{40}\,{\rm erg\,s}^{-1}$ for $\approx1$\,yr, corresponding to $\varepsilon_\nu^2dn_\nu/d\varepsilon_\nu\approx10^{48}\,{\rm erg}$, following type IIn supernovae. This extended emission is obtained assuming a $\dot{M}/v_{\rm w,7}\approx 0.1\,M_\odot$\,yr$^{-1}$ over 100\,yr preceding the explosion, corresponding to $M_{\rm CSM}=10\,M_\odot$, large ejecta energy, $\gtrsim10^{52}$\,erg, and large bolometric luminosity $L\approx10^{44}\,{\rm erg\,s}^{-1}$ lasting over a year. A possible example of such an event is SN\,2010jl \citep[e.g.,][]{Ofek+2014_NuSTAR}. However, such events are likely very rare\footnote{Assuming one 2010jl like SN was found in 15 years to a distance of 50\,Mpc over half the sky, then the rate of such events
is in the range of $5\times10^{-9}$ to $10^{-6}\,{\rm Mpc^{-3} yr^{-1}}$ (95\% CL).
}, with a small rate compared to the overall type IIn rate, i.e. $\ll 10^{-5}\,{\rm Mpc^{-3}\,yr^{-1}}$, and cannot, therefore, contribute significantly to the neutrino background (see Equation~(\ref{eq:Qnu-diff})).

\section{Discussion}
\label{sec:discussion}

The energy carried by high energy neutrinos produced during and shortly after SN shock breakout from dense CSM surrounding a super-giant progenitor is given by eqs.~(\ref{eq:Enu-log}-\ref{eq:Enu-lim}). The average neutrino production rate by the SN population is 
\begin{eqnarray}
    \label{eq:SN-nu-gen}
    (\varepsilon_\nu^2d\dot{n}_\nu/d\varepsilon_\nu)_{z=0}&=&
    4\times10^{43}\varepsilon_{\rm CR,-1}v_9^2\frac{M_{\rm CSM}}{0.1M_\odot} \nonumber \\
    &\times&\frac{R_{\rm SN}}{\rm 10^{-4}Mpc^{-3}yr^{-1}}\frac{\rm erg}{\rm Mpc^3yr},  
\end{eqnarray}
where $M_{\rm CSM}$ is the mass ejected within a few years preceding the explosion, and $R_{\rm SN}$ is the local ($z=0$) SN rate \citep{Li+2011MNRAS_NearbySN_Rates_ObservedLuminomsityFunction}. Comparing this result to Equation~(\ref{eq:Qnu-diff}), we find that if the ejection of $\approx0.1\,M_\odot$ during the years preceding a SN explosion is common, or if the average of this ejected mass over the SN population is $\approx0.1\,M_\odot$, then the resulting breakout neutrino production would account for a significant fraction of the observed background. Type IIn SNe, that occur at a rate of $\approx10^{-5}\,{\rm Mpc^{-3}\,yr^{-1}}$ and show prevalent ejection of a fraction of a solar mass preceding the explosion, may produce $\approx 0.2(M_{\rm CSM}/0.2M_\odot)$ of the background. Similarly, if an elevated mass loss, $\dot{M}/v_{\rm w,7}\approx 0.01 M_\odot$\,yr$^{-1}$, is common over a 2-year time scale preceding type II SN explosions that occur at a rate of $\approx10^{-4}\,{\rm Mpc^{-3}\,yr^{-1}}$, then these explosions would also contribute approximately 20\% of the background.

The neutrinos are expected to be produced over a few days time scale, coincident with the bright UV (possibly followed by bright X-ray) breakout emission (the collisionless shock is formed at $\approx R_{\rm bo}/2$), and to carry a fraction of $\approx0.05\varepsilon_{CR,-1}$ of the electromagnetic energy. The duration of the emission is determined by the duration $t_{\rm pr}$ of elevated mass loss preceding the explosion and given by $\approx t_{\rm pr}v_{\rm w}/v= 3(t_{\rm pr}/{1\rm yr})(v_{\rm w,7}/v_9)$\,d. 

A breakout producing $E_\nu(>1\,{\rm TeV})=10^{48}$\,erg, corresponding to $M_{\rm CSM}=0.03M_\odot$ and electromagnetic emission of $2\times10^{49}$\,erg, will produce $>1$ muon induced neutrino events (on average) in a 1\,km$^2$ detector out to a distance of $\lesssim5$\,Mpc. Type II SNe are expected to occur within this volume at a rate of $\sim1/20$\,yr. For a 10\,km$^2$ detector, the rate of SNe producing $>1$ muon-induced neutrino events would increase to $\sim1$/yr. An association by a 1\,km$^2$ detector of a single muon-induced 10\,TeV neutrino with a very nearby SN ($\lesssim10$\,Mpc) within a few days of the explosion would be significant at approximately $99.9\%$ confidence level - for an atmospheric neutrino background rate of $\approx2\times10^3/$\,yr \citep{2022IC-1068} and arrival direction uncertainty of $<1$\,squared degree, the fraction of the sky covered by the background neutrinos is $\approx10^{-3}/(3\,{\rm d})$. The current upper limits on the energy emitted in high energy neutrinos following SNe \citep[$3\times10^{48}$\,erg and $10^{49}$\,erg for SNe of type IIP and IIn respectively,][]{2023IC-SN-limit}, are larger than, but close to, the emission predicted in our analysis (see Eq.~(\ref{eq:Enu-lim})). These limits are, however, not directly applicable to the scenario that we consider: They were obtained for neutrino emission over long time windows (100~d and longer) starting at the time of first optical SN detection, which is typically later than the few days time scale over which most of the neutrino emission is predicted to occur. 

As mentioned in \S~\ref{sec:mass-loss}, the observations of SN 2023ixf are consistent with a shock breakout through a dense CSM shell, with $\dot{M}/(v_{\rm w}/100\,{\rm km{\rm\, s^{-1}}})=0.03\,M_\odot/$yr extending to $\approx2\times10^{14}$~cm, surrounded by a much lower density wind, $\dot{M}/(v_{\rm w}/100\,{\rm km{\rm\, s^{-1}}})\approx10^{-4}\,M_\odot$\,yr$^{-1}$, at larger radii \citep{2024Zimmerman-ixf}. Since the radial extent of the enhanced mass loss wind is similar in this case to the breakout radius, Equation~(\ref{eq:Rbr}), the transition from UV to X-ray emission occurs after the shock propagates into the lower density wind, producing an X-ray luminosity much lower than the UV breakout luminosity. The collisionless shock is expected to form at $R_{\rm br}/2\approx 10^{14}$\,cm, beyond which the mass of the enhanced density CSM is $\approx10^{-2}\,M_\odot$. We, therefore, expect $E_\nu(>1\,{\rm TeV})\approx2\times 10^{47}$\,erg (see Equation~(\ref{eq:Enu-log})) corresponding to $0.2$ muon induced neutrino events in a 1\,km$^2$ detector for a distance of $6.4$\,Mpc, consistent with the IceCube non-detection \citep{2023ATel16043_SN2023ixf_IceCubeLimits}. The high energy, $>1$\,GeV, gamma-ray luminosity expected to accompany the collisionless shock, approximately $2/3$ of the neutrino luminosity, is $\approx10^{42}\,{\rm erg\,s}^{-1}$ over $\approx1\,$d. This is in some tension with the {\it Fermi} 95\% confidence upper limit on the (1\,d duration) lumnosity of $\approx 0.5\times 10^{42}\,{\rm erg\,s}^{-1}$ \citep{Marti-2024A&A_SN2023ixf_FermiObservations}. This may indicate that the fraction $\varepsilon_{\rm CR}$ of the shock thermal energy carried by non-thermal relativistic protons is somewhat smaller than the 0.1 value we adopted. However, given the uncertainty in the wind and shock parameter estimates, the fact that the enhanced mass-loss wind terminates in this case near the breakout radius, and the lack of a complete quantitative theory of wind shock breakout, an accurate quantitative conclusion regarding the value of $\varepsilon_{\rm CR}$ cannot yet be drawn.  

The neutrino luminosity predicted in \citet{2024Murase-SNe} for SN 2003ixf is significantly lower than predicted here mainly because they adopt CSM density and shock velocity that yield a substantially lower rate of shock energy deposition in the CSM, $7\times10^{42}\,{\rm erg\,s}^{-1}$ compared to our $\approx10^{44}\,{\rm erg\,s}^{-1}$ (see Eq.~(\ref{eq:Lbr}); the discrepancy in total fluence is smaller). Our parameter choice is based on observations, including in particular the observed $\approx3\times10^{43}\,{\rm erg\,s}^{-1}$ luminosity, that were not available at the time of their paper submission. When similar values are adopted for the CSM density and shock velocity (and CR acceleration efficiency), our neutrino luminosity estimate is similar to what would be obtained by \citet{2024Murase-SNe}.

While our discussion has focused on type II SNe, which is the most common type of events, recent evidence shows that other types of massive star explosions also occur within compact CSM, including some type Ic events lacking hydrogen and helium \citep{irani2024_oqm}, as well as rare populations of events where the CSM is dominated by helium \citep[Type Ibn,][]{Pastorello+2007_SN2006jc_precursor,Pastorello08, Karamehmetoglu21,Hosseinzadeh+2017_SN_Ibn_diversity}, or carbon and oxygen \citep[type Icn,][]{Gal-Yam+2022Natur_AT2019hgp_Icn_SN,Perley22,Pellegrino22}. The equations derived in \S~\ref{sec:BO} for the neutrino emission are valid for all progenitors surrounded by hydrogen-dominated CSM and are straightforward to generalize to He- and C/O-dominated CSM. While the rate of type Ibc SNe is $\approx1/3$ the type II SN rate, their contribution to the detected cosmic high-energy neutrino signal may be significant if compact dense CSM is common for such explosions, noting particularly that they may be characterized by higher breakout velocities due to the more compact progenitor structure. Further observations are required to constrain the CSM properties of this population.

As the capabilities of rapid transient searches improve, both from the ground \citep[e.g.][]{GOTO2022,Ofek+2023PASP_LAST_Overview} and from space with the expected launch of the wide-field UV space telescope ULTRASAT \citep{2024Shvartzvald-ULTRASAT}, a systematic detection of many SNe of all types at early, $<1$\,d, time will be possible. Early UV measurements, early spectra, and possibly early X-ray measurements will allow us to determine the properties of mass loss over years times scale preceding the explosion for the SN progenitor population. 

The derivation of quantitative constraints on the mass loss history and pre-explosion envelope structure, and a quantitative determination of the neutrino luminosity and spectrum, will require a quantitative theory describing the evolution of the electromagnetic spectrum during a breakout as the radiation-mediated shock transforms into a collisionless one. Theoretical analyses are challenged by the non-steady nature of the problem (with shock structure evolving over a dynamical time, rendering steady-state solutions inapplicable and greatly complicating analytic analyses), by the deviation from an equilibrium between the electrons and protons, and by the importance of inelastic Compton scattering in determining the (quasi thermal) electron energy temperature and the optical-X-ray spectrum. No current analysis \citep[e.g.][]{ChevalierIrwin12,Moriya13,SvirskiNakar14,GinzburgBalberg14,KasenNumeric15,HayniePiro21CSMBO,MargalitThickCSM22} addresses all these issues self-consistently (see Wasserman et al. 2024 for discussion). While the formation of a non-thermal high-energy particle population by shock acceleration is not expected to modify significantly the shock dynamics and UV-X-ray emission (KSW11, Wasserman et al. 2024), the background UV-X-ray spectrum strongly affects the production of high-energy radiation. As noted above, for example, the UV intensity limits the acceleration of protons, and the escape of 1\,GeV photons depends on the interplay between collisionless shock formation and the spectrum shift to the X-ray band.

\begin{acknowledgments}
We thank K. Murase for useful discussions and the anonymous referee for significant constructive comments. E.W.'s research is partially supported by grants from the André Deloro Institute for Space and Optics Research, the Norman E Alexander Family M Foundation ULTRASAT Data Center Fund, and the Sagol Weizmann-MIT Bridge Program. E.O.O. is grateful for the support of grants from the Norman E Alexander Family M Foundation ULTRASAT Data Center Fund, and the Willner Family Leadership Institute for the Weizmann Institute of Science, Willner Family Leadership Institute, André Deloro Institute, Paul and Tina Gardner, Israel Science Foundation, NSF-BSF, Israel Council for Higher Education (VATAT), Sagol Weizmann-MIT, and the Rosa and Emilio Segre Research Award. AGY’s research is supported by the André Deloro Institute for Space and Optics Research, the Center for Experimental Physics, a WIS-MIT Sagol grant, the Norman E Alexander Family M Foundation ULTRASAT Data Center Fund, and Yeda-Sela; AGY is the incumbent of the The Arlyn Imberman Professorial Chair.
\end{acknowledgments}

\bibliographystyle{aasjournal}

\begin{thebibliography}{}
\expandafter\ifx\csname natexlab\endcsname\relax\def\natexlab#1{#1}\fi
\providecommand{\url}[1]{\href{#1}{#1}}
\providecommand{\dodoi}[1]{doi:~\href{http://doi.org/#1}{\nolinkurl{#1}}}
\providecommand{\doeprint}[1]{\href{http://ascl.net/#1}{\nolinkurl{http://ascl.net/#1}}}
\providecommand{\doarXiv}[1]{\href{https://arxiv.org/abs/#1}{\nolinkurl{https://arxiv.org/abs/#1}}}

\bibitem[{{Aartsen} {et~al.}(2017){Aartsen}, {Abraham}, {Ackermann}, {Adams}, {Aguilar}, {Ahlers}, {Ahrens}, {Altmann}, {Andeen}, {Anderson}, {Ansseau}, {Anton}, {Archinger}, {Arguelles}, {Arlen}, {Auffenberg}, {Axani}, {Bai}, {Barwick}, {Baum}, {Bay}, {Beatty}, {Becker Tjus}, {Becker}, {BenZvi}, {Berghaus}, {Berley}, {Bernardini}, {Bernhard}, {Besson}, {Binder}, {Bindig}, {Bissok}, {Blaufuss}, {Blot}, {Boersma}, {Bohm}, {B{\"o}rner}, {Bos}, {Bose}, {B{\"o}ser}, {Botner}, {Braun}, {Brayeur}, {Bretz}, {Burgman}, {Casey}, {Casier}, {Cheung}, {Chirkin}, {Christov}, {Clark}, {Classen}, {Coenders}, {Collin}, {Conrad}, {Cowen}, {Cruz Silva}, {Daughhetee}, {Davis}, {Day}, {de Andr{\'e}}, {De Clercq}, {del Pino Rosendo}, {Dembinski}, {De Ridder}, {Desiati}, {de Vries}, {de Wasseige}, {de With}, {DeYoung}, {D{\'\i}az-V{\'e}lez}, {di Lorenzo}, {Dujmovic}, {Dumm}, {Dunkman}, {Eberhardt}, {Ehrhardt}, {Eichmann}, {Euler}, {Evenson}, {Fahey}, {Fazely}, {Feintzeig}, {Felde}, {Filimonov}, {Finley}, {Flis}, {F{\"o}sig},
  {Franckowiak}, {Fuchs}, {Gaisser}, {Gaior}, {Gallagher}, {Gerhardt}, {Ghorbani}, {Giang}, {Gladstone}, {Glagla}, {Gl{\"u}senkamp}, {Goldschmidt}, {Golup}, {Gonzalez}, {G{\'o}ra}, {Grant}, {Griffith}, {Haack}, {Haj Ismail}, {Hallgren}, {Halzen}, {Hansen}, {Hansmann}, {Hansmann}, {Hanson}, {Hebecker}, {Heereman}, {Helbing}, {Hellauer}, {Hickford}, {Hignight}, {Hill}, {Hoffman}, {Hoffmann}, {Holzapfel}, {Homeier}, {Hoshina}, {Huang}, {Huber}, {Huelsnitz}, {Hultqvist}, {In}, {Ishihara}, {Jacobi}, {Japaridze}, {Jeong}, {Jero}, {Jones}, {Jurkovic}, {Kappes}, {Karg}, {Karle}, {Katz}, {Kauer}, {Keivani}, {Kelley}, {Kemp}, {Kheirandish}, {Kim}, {Kintscher}, {Kiryluk}, {Kittler}, {Klein}, {Kohnen}, {Koirala}, {Kolanoski}, {Konietz}, {K{\"o}pke}, {Kopper}, {Kopper}, {Koskinen}, {Kowalski}, {Krings}, {Kroll}, {Kr{\"u}ckl}, {Kr{\"u}ger}, {Kunnen}, {Kunwar}, {Kurahashi}, {Kuwabara}, {Labare}, {Lanfranchi}, {Larson}, {Lennarz}, {Lesiak-Bzdak}, {Leuermann}, {Leuner}, {Lu}, {L{\"u}nemann}, {Madsen}, {Maggi}, {Mahn},
  {Mancina}, {Mandelartz}, {Maruyama}, {Mase}, {Maunu}, {McNally}, {Meagher}, {Medici}, {Meier}, {Meli}, {Menne}, {Merino}, {Meures}, {Miarecki}, {Middell}, {Mohrmann}, {Montaruli}, {Moulai}, {Nahnhauer}, {Naumann}, {Neer}, {Niederhausen}, {Nowicki}, {Nygren}, {Obertacke Pollmann}, {Olivas}, {Omairat}, {O'Murchadha}, {Palczewski}, {Pandya}, {Pankova}, {Penek}, {Pepper}, {P{\'e}rez de los Heros}, {Pfendner}, {Pieloth}, {Pinat}, {Posselt}, {Price}, {Przybylski}, {Quinnan}, {Raab}, {R{\"a}del}, {Rameez}, {Rawlins}, {Reimann}, {Relich}, {Resconi}, {Rhode}, {Richman}, {Riedel}, {Robertson}, {Rongen}, {Rott}, {Ruhe}, {Ryckbosch}, {Rysewyk}, {Sabbatini}, {Sanchez Herrera}, {Sandrock}, {Sandroos}, {Sarkar}, {Satalecka}, {Schimp}, {Schlunder}, {Schmidt}, {Schoenen}, {Sch{\"o}neberg}, {Sch{\"o}nwald}, {Schumacher}, {Seckel}, {Seunarine}, {Soldin}, {Song}, {Spiczak}, {Spiering}, {Stahlberg}, {Stamatikos}, {Stanev}, {Stasik}, {Steuer}, {Stezelberger}, {Stokstad}, {St{\"o}{\ss}l}, {Str{\"o}m}, {Strotjohann}, {Sullivan},
  {Sutherland}, {Taavola}, {Taboada}, {Tatar}, {Ter-Antonyan}, {Terliuk}, {Te{\v{s}}i{\'c}}, {Tilav}, {Toale}, {Tobin}, {Toscano}, {Tosi}, {Tselengidou}, {Turcati}, {Unger}, {Usner}, {Vallecorsa}, {Vandenbroucke}, {van Eijndhoven}, {Vanheule}, {van Rossem}, {van Santen}, {Veenkamp}, {Vehring}, {Voge}, {Vraeghe}, {Walck}, {Wallace}, {Wallraff}, {Wandkowsky}, {Weaver}, {Wendt}, {Westerhoff}, {Whelan}, {Wickmann}, {Wiebe}, {Wiebusch}, {Wille}, {Williams}, {Wills}, {Wissing}, {Wolf}, {Wood}, {Woolsey}, {Woschnagg}, {Xu}, {Xu}, {Xu}, {Yanez}, {Yodh}, {Yoshida}, {Zoll}, \& {IceCube Collaboration}}]{2017IC-blazar-limit}
{Aartsen}, M.~G., {Abraham}, K., {Ackermann}, M., {et~al.} 2017, \apj, 835, 45, \dodoi{10.3847/1538-4357/835/1/45}

\bibitem[{{Abbasi} {et~al.}(2023){Abbasi}, {Ackermann}, {Adams}, {Agarwalla}, {Aguilar}, {Ahlers}, {Alameddine}, {Amin}, {Andeen}, {Anton}, {Arg{\"u}elles}, {Ashida}, {Athanasiadou}, {Axani}, {Bai}, {Balagopal}, {Baricevic}, {Barwick}, {Basu}, {Bay}, {Beatty}, {Becker}, {Becker Tjus}, {Beise}, {Bellenghi}, {BenZvi}, {Berley}, {Bernardini}, {Besson}, {Binder}, {Bindig}, {Blaufuss}, {Blot}, {Bontempo}, {Book}, {Meneguolo}, {B{\"o}ser}, {Botner}, {B{\"o}ttcher}, {Bourbeau}, {Braun}, {Brinson}, {Brostean-Kaiser}, {Burley}, {Busse}, {Butterfield}, {Campana}, {Carloni}, {Carnie-Bronca}, {Chattopadhyay}, {Chen}, {Chen}, {Chirkin}, {Choi}, {Clark}, {Classen}, {Coleman}, {Collin}, {Connolly}, {Conrad}, {Coppin}, {Correa}, {Countryman}, {Cowen}, {Dave}, {De Clercq}, {DeLaunay}, {Delgado L{\'o}pez}, {Dembinski}, {Deoskar}, {Desai}, {Desiati}, {de Vries}, {de Wasseige}, {DeYoung}, {Diaz}, {D{\'\i}az-V{\'e}lez}, {Dittmer}, {Domi}, {Dujmovic}, {DuVernois}, {Ehrhardt}, {Eller}, {Engel}, {Erpenbeck}, {Evans}, {Evenson},
  {Fan}, {Fang}, {Fazely}, {Fedynitch}, {Feigl}, {Fiedlschuster}, {Finley}, {Fischer}, {Fox}, {Franckowiak}, {Friedman}, {Fritz}, {F{\"u}rst}, {Gaisser}, {Gallagher}, {Ganster}, {Garcia}, {Garrappa}, {Gerhardt}, {Ghadimi}, {Glaser}, {Glauch}, {Gl{\"u}senkamp}, {Goehlke}, {Gonzalez}, {Goswami}, {Grant}, {Gray}, {Griffin}, {Griswold}, {G{\"u}nther}, {Gutjahr}, {Haack}, {Hallgren}, {Halliday}, {Halve}, {Halzen}, {Hamdaoui}, {Ha Minh}, {Hanson}, {Hardin}, {Harnisch}, {Hatch}, {Haungs}, {Helbing}, {Hellrung}, {Henningsen}, {Heuermann}, {Hickford}, {Hidvegi}, {Hill}, {Hill}, {Hoffman}, {Hoshina}, {Hou}, {Huber}, {Hultqvist}, {H{\"u}nnefeld}, {Hussain}, {Hymon}, {In}, {Iovine}, {Ishihara}, {Jacquart}, {Jansson}, {Japaridze}, {Jayakumar}, {Jeong}, {Jin}, {Jones}, {Kang}, {Kang}, {Kang}, {Kappes}, {Kappesser}, {Kardum}, {Karg}, {Karl}, {Karle}, {Katz}, {Kauer}, {Kelley}, {Khatee Zathul}, {Kheirandish}, {Kin}, {Kiryluk}, {Klein}, {Kochocki}, {Koirala}, {Kolanoski}, {Kontrimas}, {K{\"o}pke}, {Kopper}, {Koskinen},
  {Koundal}, {Kovacevich}, {Kowalski}, {Kozynets}, {Kruiswijk}, {Krupczak}, {Kumar}, {Kun}, {Kurahashi}, {Lad}, {Lagunas Gualda}, {Lamoureux}, {Larson}, {Lauber}, {Lazar}, {Lee}, {Leonard DeHolton}, {Leszczy{\'n}ska}, {Lincetto}, {Liu}, {Liubarska}, {Lohfink}, {Love}, {Mariscal}, {Lu}, {Lucarelli}, {Ludwig}, {Luszczak}, {Lyu}, {Ma}, {Madsen}, {Mahn}, {Makino}, {Mancina}, {Marie Sainte}, {Mari{\c{s}}}, {Marka}, {Marka}, {Marsee}, {Martinez-Soler}, {Maruyama}, {Mayhew}, {McElroy}, {McNally}, {Mead}, {Meagher}, {Mechbal}, {Medina}, {Meier}, {Meighen-Berger}, {Merckx}, {Merten}, {Micallef}, {Mockler}, {Montaruli}, {Moore}, {Morii}, {Morse}, {Moulai}, {Mukherjee}, {Naab}, {Nagai}, {Nakos}, {Naumann}, {Necker}, {Neumann}, {Niederhausen}, {Nisa}, {Noell}, {Nowicki}, {Obertacke Pollmann}, {O'Dell}, {Oehler}, {Oeyen}, {Olivas}, {Orsoe}, {Osborn}, {O'Sullivan}, {Pandya}, {Park}, {Parker}, {Paudel}, {Paul}, {P{\'e}rez de los Heros}, {Peterson}, {Philippen}, {Pieper}, {Pizzuto}, {Plum}, {Popovych}, {Prado Rodriguez},
  {Pries}, {Procter-Murphy}, {Przybylski}, {Raab}, {Rack-Helleis}, {Rawlins}, {Rechav}, {Rehman}, {Reichherzer}, {Renzi}, {Resconi}, {Reusch}, {Rhode}, {Richman}, {Riedel}, {Roberts}, {Robertson}, {Rodan}, {Roellinghoff}, {Rongen}, {Rott}, {Ruhe}, {Ruohan}, {Ryckbosch}, {Safa}, {Saffer}, {Salazar-Gallegos}, {Sampathkumar}, {Sanchez Herrera}, {Sandrock}, {Santander}, {Sarkar}, {Sarkar}, {Savelberg}, {Savina}, {Schaufel}, {Schieler}, {Schindler}, {Schl{\"u}ter}, {Schmidt}, {Schneider}, {Schr{\"o}der}, {Schumacher}, {Schwefer}, {Sclafani}, {Seckel}, {Seunarine}, {Sharma}, {Shefali}, {Shimizu}, {Silva}, {Skrzypek}, {Smithers}, {Snihur}, {Soedingrekso}, {S{\o}gaard}, {Soldin}, {Sommani}, {Spannfellner}, {Spiczak}, {Spiering}, {Stamatikos}, {Stanev}, {Stasik}, {Stein}, {Stezelberger}, {St{\"u}rwald}, {Stuttard}, {Sullivan}, {Taboada}, {Ter-Antonyan}, {Thompson}, {Thwaites}, {Tilav}, {Tollefson}, {T{\"o}nnis}, {Toscano}, {Tosi}, {Trettin}, {Tung}, {Turcotte}, {Twagirayezu}, {Ty}, {Unland Elorrieta}, {Upadhyay},
  {Upshaw}, {Valtonen-Mattila}, {Vandenbroucke}, {van Eijndhoven}, {Vannerom}, {van Santen}, {Vara}, {Veitch-Michaelis}, {Venugopal}, {Verpoest}, {Veske}, {Walck}, {Watson}, {Weaver}, {Weigel}, {Weindl}, {Weldert}, {Wendt}, {Werthebach}, {Weyrauch}, {Whitehorn}, {Wiebusch}, {Willey}, {Williams}, {Wolf}, {Wrede}, {Wulff}, {Xu}, {Yanez}, {Yildizci}, {Yoshida}, {Yu}, {Yu}, {Yuan}, {Zhang}, {Zhelnin}, \& {IceCube Collaboration}}]{2023IC-SN-limit}
{Abbasi}, R., {Ackermann}, M., {Adams}, J., {et~al.} 2023, \apjl, 949, L12, \dodoi{10.3847/2041-8213/acd2c9}

\bibitem[{{Ackermann} {et~al.}(2015){Ackermann}, {Ajello}, {Albert}, {Atwood}, {Baldini}, {Ballet}, {Barbiellini}, {Bastieri}, {Bechtol}, {Bellazzini}, {Bissaldi}, {Blandford}, {Bloom}, {Bottacini}, {Brandt}, {Bregeon}, {Bruel}, {Buehler}, {Buson}, {Caliandro}, {Cameron}, {Caragiulo}, {Caraveo}, {Cavazzuti}, {Cecchi}, {Charles}, {Chekhtman}, {Chiang}, {Chiaro}, {Ciprini}, {Claus}, {Cohen-Tanugi}, {Conrad}, {Cuoco}, {Cutini}, {D'Ammando}, {de Angelis}, {de Palma}, {Dermer}, {Digel}, {Silva}, {Drell}, {Favuzzi}, {Ferrara}, {Focke}, {Franckowiak}, {Fukazawa}, {Funk}, {Fusco}, {Gargano}, {Gasparrini}, {Germani}, {Giglietto}, {Giommi}, {Giordano}, {Giroletti}, {Godfrey}, {Gomez-Vargas}, {Grenier}, {Guiriec}, {Gustafsson}, {Hadasch}, {Hayashi}, {Hays}, {Hewitt}, {Ippoliti}, {Jogler}, {J{\'o}hannesson}, {Johnson}, {Johnson}, {Kamae}, {Kataoka}, {Kn{\"o}dlseder}, {Kuss}, {Larsson}, {Latronico}, {Li}, {Li}, {Longo}, {Loparco}, {Lott}, {Lovellette}, {Lubrano}, {Madejski}, {Manfreda}, {Massaro}, {Mayer}, {Mazziotta},
  {McEnery}, {Michelson}, {Mitthumsiri}, {Mizuno}, {Moiseev}, {Monzani}, {Morselli}, {Moskalenko}, {Murgia}, {Nemmen}, {Nuss}, {Ohsugi}, {Omodei}, {Orlando}, {Ormes}, {Paneque}, {Panetta}, {Perkins}, {Pesce-Rollins}, {Piron}, {Pivato}, {Porter}, {Rain{\`o}}, {Rando}, {Razzano}, {Razzaque}, {Reimer}, {Reimer}, {Reposeur}, {Ritz}, {Romani}, {S{\'a}nchez-Conde}, {Schaal}, {Schulz}, {Sgr{\`o}}, {Siskind}, {Spandre}, {Spinelli}, {Strong}, {Suson}, {Takahashi}, {Thayer}, {Thayer}, {Tibaldo}, {Tinivella}, {Torres}, {Tosti}, {Troja}, {Uchiyama}, {Vianello}, {Werner}, {Winer}, {Wood}, {Wood}, {Zaharijas}, \& {Zimmer}}]{2015Fermi-XGB}
{Ackermann}, M., {Ajello}, M., {Albert}, A., {et~al.} 2015, \apj, 799, 86, \dodoi{10.1088/0004-637X/799/1/86}

\bibitem[{{Ackermann} {et~al.}(2016){Ackermann}, {Ajello}, {Albert}, {Atwood}, {Baldini}, {Ballet}, {Barbiellini}, {Bastieri}, {Bechtol}, {Bellazzini}, {Bissaldi}, {Blandford}, {Bloom}, {Bonino}, {Bregeon}, {Britto}, {Bruel}, {Buehler}, {Caliandro}, {Cameron}, {Caragiulo}, {Caraveo}, {Cavazzuti}, {Cecchi}, {Charles}, {Chekhtman}, {Chiang}, {Chiaro}, {Ciprini}, {Cohen-Tanugi}, {Cominsky}, {Costanza}, {Cutini}, {D'Ammando}, {de Angelis}, {de Palma}, {Desiante}, {Digel}, {Di Mauro}, {Di Venere}, {Dom{\'\i}nguez}, {Drell}, {Favuzzi}, {Fegan}, {Ferrara}, {Franckowiak}, {Fukazawa}, {Funk}, {Fusco}, {Gargano}, {Gasparrini}, {Giglietto}, {Giommi}, {Giordano}, {Giroletti}, {Godfrey}, {Green}, {Grenier}, {Guiriec}, {Hays}, {Horan}, {Iafrate}, {Jogler}, {J{\'o}hannesson}, {Kuss}, {La Mura}, {Larsson}, {Latronico}, {Li}, {Li}, {Longo}, {Loparco}, {Lott}, {Lovellette}, {Lubrano}, {Madejski}, {Magill}, {Maldera}, {Manfreda}, {Mayer}, {Mazziotta}, {Michelson}, {Mitthumsiri}, {Mizuno}, {Moiseev}, {Monzani}, {Morselli},
  {Moskalenko}, {Murgia}, {Negro}, {Nuss}, {Ohsugi}, {Okada}, {Omodei}, {Orlando}, {Ormes}, {Paneque}, {Perkins}, {Pesce-Rollins}, {Petrosian}, {Piron}, {Pivato}, {Porter}, {Rain{\`o}}, {Rando}, {Razzano}, {Razzaque}, {Reimer}, {Reimer}, {Reposeur}, {Romani}, {S{\'a}nchez-Conde}, {Schmid}, {Schulz}, {Sgr{\`o}}, {Simone}, {Siskind}, {Spada}, {Spandre}, {Spinelli}, {Suson}, {Takahashi}, {Thayer}, {Tibaldo}, {Torres}, {Troja}, {Vianello}, {Yassine}, \& {Zimmer}}]{2016Fermi-gbgnd-blazars}
---. 2016, \prl, 116, 151105, \dodoi{10.1103/PhysRevLett.116.151105}

\bibitem[{{Ackermann} {et~al.}(2022){Ackermann}, {Bustamante}, {Lu}, {Otte}, {Reno}, {Wissel}, {Ackermann}, {Agarwalla}, {Alvarez-Mu{\~n}iz}, {Alves Batista}, {Arg{\"u}elles}, {Bustamante}, {Clark}, {Cummings}, {Das}, {Decoene}, {Denton}, {Dornic}, {Dzhilkibaev}, {Farzan}, {Garcia}, {Garzelli}, {Glaser}, {Heijboer}, {H{\"o}randel}, {Illuminati}, {Seon Jeong}, {Kelley}, {Kelly}, {Kheirandish}, {Klein}, {Krizmanic}, {Larson}, {Lu}, {Murase}, {Narang}, {Otte}, {Prechelt}, {Prohira}, {Reno}, {Resconi}, {Santander}, {Valera}, {Vandenbroucke}, {Vasil'evna Suvorova}, {Wiencke}, {Wissel}, {Yoshida}, {Yuan}, {Zas}, {Zhelnin}, {Zhou}, {Anchordoqui}, {Ashida}, {Bagheri}, {Balagopal}, {Basu}, {Beatty}, {Bechtol}, {Bell}, {Bishop}, {Book}, {Brown}, {Burgman}, {Campana}, {Chau}, {Chen}, {Coleman}, {Connolly}, {Conrad}, {Correa}, {Creque-Sarbinowski}, {Cummings}, {Curtis-Ginsberg}, {Dasgupta}, {De Kockere}, {de Vries}, {Deaconu}, {Desai}, {DeYoung}, {di Matteo}, {Elsaesser}, {F{\"u}rst}, {Fan}, {Fedynitch}, {Fox},
  {Ganster}, {Minh}, {Haack}, {Hallman}, {Halzen}, {Haungs}, {Ishihara}, {Judd}, {Karg}, {Karle}, {Katori}, {Kochocki}, {Kopper}, {Kowalski}, {Kravchenko}, {Kurahashi}, {Lamoureux}, {Le{\'o}n Vargas}, {Lincetto}, {Liu}, {Madsen}, {Makino}, {Mammo}, {Marka}, {Mayotte}, {Meagher}, {Meier}, {Minh}, {Miramonti}, {Moulai}, {Mulrey}, {Muzio}, {Naab}, {Nelles}, {Nichols}, {Nozdrina}, {O'Sullivan}, {OD́ell}, {Osborne}, {Pandey}, {Paudel}, {Pizzuto}, {Plum}, {Pobes Aranda}, {Pyras}, {Raab}, {Rechav}, {Rojo}, {Romero Matamala}, {Santander}, {Savina}, {Schroeder}, {Schumacher}, {Sciutto}, {Sclafani}, {Ful Hossain Seikh}, {Silva}, {Singh}, {Smith}, {Spencer}, {Springer}, {Stachurska}, {Suvorova}, {Taboada}, {Toscano}, {Tueros}, {Twagirayezu}, {van Eijndhoven}, {Veres}, {Vieregg}, {Wang}, {Whitehorn}, {Winter}, {Yildizci}, \& {Yu}}]{2022Snowmass-nu}
{Ackermann}, M., {Bustamante}, M., {Lu}, L., {et~al.} 2022, Journal of High Energy Astrophysics, 36, 55, \dodoi{10.1016/j.jheap.2022.08.001}

\bibitem[{{Ando} {et~al.}(2015){Ando}, {Tamborra}, \& {Zandanel}}]{2015Ando-g-constraint}
{Ando}, S., {Tamborra}, I., \& {Zandanel}, F. 2015, \prl, 115, 221101, \dodoi{10.1103/PhysRevLett.115.221101}

\bibitem[{{Blandford} \& {Eichler}(1987)}]{1987BlandfordEichler}
{Blandford}, R., \& {Eichler}, D. 1987, \physrep, 154, 1, \dodoi{10.1016/0370-1573(87)90134-7}

\bibitem[{{Boian} \& {Groh}(2019)}]{2019Boian-flash-wind}
{Boian}, I., \& {Groh}, J.~H. 2019, \aap, 621, A109, \dodoi{10.1051/0004-6361/201833779}

\bibitem[{{Bostroem} {et~al.}(2023){Bostroem}, {Pearson}, {Shrestha}, {Sand}, {Valenti}, {Jha}, {Andrews}, {Smith}, {Terreran}, {Green}, {Dong}, {Lundquist}, {Haislip}, {Hoang}, {Hosseinzadeh}, {Janzen}, {Jencson}, {Kouprianov}, {Paraskeva}, {Meza Retamal}, {Reichart}, {Arcavi}, {Bonanos}, {Coughlin}, {Dobson}, {Farah}, {Galbany}, {Guti{\'e}rrez}, {Hawley}, {Hebb}, {Hiramatsu}, {Howell}, {Iijima}, {Ilyin}, {Jhass}, {McCully}, {Moran}, {Morris}, {Mura}, {M{\"u}ller-Bravo}, {Munday}, {Newsome}, {Pabst}, {Ochner}, {Gonzalez}, {Pastorello}, {Pellegrino}, {Piscarreta}, {Ravi}, {Reguitti}, {Salo}, {Vink{\'o}}, {de Vos}, {Wheeler}, {Williams}, \& {Wyatt}}]{2023Bostroem-ixf}
{Bostroem}, K.~A., {Pearson}, J., {Shrestha}, M., {et~al.} 2023, \apjl, 956, L5, \dodoi{10.3847/2041-8213/acf9a4}

\bibitem[{{Bruch} {et~al.}(2021){Bruch}, {Gal-Yam}, {Schulze}, {Yaron}, {Yang}, {Soumagnac}, {Rigault}, {Strotjohann}, {Ofek}, {Sollerman}, {Masci}, {Barbarino}, {Ho}, {Fremling}, {Perley}, {Nordin}, {Cenko}, {Adams}, {Adreoni}, {Bellm}, {Blagorodnova}, {Bulla}, {Burdge}, {De}, {Dhawan}, {Drake}, {Duev}, {Dugas}, {Graham}, {Graham}, {Irani}, {Jencson}, {Karamehmetoglu}, {Kasliwal}, {Kim}, {Kulkarni}, {Kupfer}, {Liang}, {Mahabal}, {Miller}, {Prince}, {Riddle}, {Sharma}, {Smith}, {Taddia}, {Taggart}, {Walters}, \& {Yan}}]{Bruch+2021_SN_Progenitors_ElevatedMassLoss_FlashSpectroscopy}
{Bruch}, R.~J., {Gal-Yam}, A., {Schulze}, S., {et~al.} 2021, \apj, 912, 46, \dodoi{10.3847/1538-4357/abef05}

\bibitem[{{Bruch} {et~al.}(2023){Bruch}, {Gal-Yam}, {Yaron}, {Chen}, {Strotjohann}, {Irani}, {Zimmerman}, {Schulze}, {Yang}, {Kim}, {Bulla}, {Sollerman}, {Rigault}, {Ofek}, {Soumagnac}, {Masci}, {Fremling}, {Perley}, {Nordin}, {Cenko}, {Ho}, {Adams}, {Adreoni}, {Bellm}, {Blagorodnova}, {Burdge}, {De}, {Dekany}, {Dhawan}, {Drake}, {Duev}, {Graham}, {Graham}, {Jencson}, {Karamehmetoglu}, {Kasliwal}, {Kulkarni}, {Miller}, {Neill}, {Prince}, {Riddle}, {Rusholme}, {Sharma}, {Smith}, {Sravan}, {Taggart}, {Walters}, \& {Yan}}]{Bruch2023}
{Bruch}, R.~J., {Gal-Yam}, A., {Yaron}, O., {et~al.} 2023, \apj, 952, 119, \dodoi{10.3847/1538-4357/acd8be}

\bibitem[{{Chevalier}(2012)}]{2012Chevalier-binary-Mloss}
{Chevalier}, R.~A. 2012, \apjl, 752, L2, \dodoi{10.1088/2041-8205/752/1/L2}

\bibitem[{{Chevalier} \& {Irwin}(2012)}]{ChevalierIrwin12}
{Chevalier}, R.~A., \& {Irwin}, C.~M. 2012, \apjl, 747, L17, \dodoi{10.1088/2041-8205/747/1/L17}

\bibitem[{{Das} {et~al.}(2024){Das}, {Zhang}, \& {Murase}}]{2024Murase-1068}
{Das}, A., {Zhang}, B.~T., \& {Murase}, K. 2024, \apj, 972, 44, \dodoi{10.3847/1538-4357/ad5a04}

\bibitem[{{de Jager} {et~al.}(1988){de Jager}, {Nieuwenhuijzen}, \& {van der Hucht}}]{1988Jager-Mdot}
{de Jager}, C., {Nieuwenhuijzen}, H., \& {van der Hucht}, K.~A. 1988, \aaps, 72, 259

\bibitem[{{Denton} \& {Tamborra}(2018)}]{2018Tamborra-chocked}
{Denton}, P.~B., \& {Tamborra}, I. 2018, \jcap, 2018, 058, \dodoi{10.1088/1475-7516/2018/04/058}

\bibitem[{{Dessart} {et~al.}(2017){Dessart}, {Hillier}, \& {Audit}}]{2017Dessart-flash-wind}
{Dessart}, L., {Hillier}, D.~J., \& {Audit}, E. 2017, \aap, 605, A83, \dodoi{10.1051/0004-6361/201730942}

\bibitem[{{F{\"o}rster} {et~al.}(2018){F{\"o}rster}, {Moriya}, {Maureira}, {Anderson}, {Blinnikov}, {Bufano}, {Cabrera-Vives}, {Clocchiatti}, {de Jaeger}, {Est{\'e}vez}, {Galbany}, {Gonz{\'a}lez-Gait{\'a}n}, {Gr{\"a}fener}, {Hamuy}, {Hsiao}, {Huentelemu}, {Huijse}, {Kuncarayakti}, {Mart{\'\i}nez}, {Medina}, {Olivares E.}, {Pignata}, {Razza}, {Reyes}, {San Mart{\'\i}n}, {Smith}, {Vera}, {Vivas}, {de Ugarte Postigo}, {Yoon}, {Ashall}, {Fraser}, {Gal-Yam}, {Kankare}, {Le Guillou}, {Mazzali}, {Walton}, \& {Young}}]{Forster2018}
{F{\"o}rster}, F., {Moriya}, T.~J., {Maureira}, J.~C., {et~al.} 2018, Nature Astronomy, 2, 808, \dodoi{10.1038/s41550-018-0563-4}

\bibitem[{{Fuller}(2017)}]{Fuller2017_precursors_RSG}
{Fuller}, J. 2017, \mnras, 470, 1642, \dodoi{10.1093/mnras/stx1314}

\bibitem[{{Fuller} \& {Ro}(2018)}]{Fuller+2018_precursors}
{Fuller}, J., \& {Ro}, S. 2018, \mnras, 476, 1853, \dodoi{10.1093/mnras/sty369}

\bibitem[{{Fusco} \& {Versari}(2019)}]{2019ANTARES-diffuse}
{Fusco}, L.~A., \& {Versari}, F. 2019, in International Cosmic Ray Conference, Vol.~36, 36th International Cosmic Ray Conference (ICRC2019), 891, \dodoi{10.22323/1.358.0891}

\bibitem[{{Gal-Yam}(2017)}]{Gal-Yam2017}
{Gal-Yam}, A. 2017, in Handbook of Supernovae, ed. A.~W. {Alsabti} \& P.~{Murdin}, 195, \dodoi{10.1007/978-3-319-21846-5_35}

\bibitem[{{Gal-Yam} {et~al.}(2014){Gal-Yam}, {Arcavi}, {Ofek}, {Ben-Ami}, {Cenko}, {Kasliwal}, {Cao}, {Yaron}, {Tal}, {Silverman}, {Horesh}, {De Cia}, {Taddia}, {Sollerman}, {Perley}, {Vreeswijk}, {Kulkarni}, {Nugent}, {Filippenko}, \& {Wheeler}}]{Gal-Yam+2014_SN2013cu_FlashSpectroscopy}
{Gal-Yam}, A., {Arcavi}, I., {Ofek}, E.~O., {et~al.} 2014, \nat, 509, 471, \dodoi{10.1038/nature13304}

\bibitem[{{Gal-Yam} {et~al.}(2022){Gal-Yam}, {Bruch}, {Schulze}, {Yang}, {Perley}, {Irani}, {Sollerman}, {Kool}, {Soumagnac}, {Yaron}, {Strotjohann}, {Zimmerman}, {Barbarino}, {Kulkarni}, {Kasliwal}, {De}, {Yao}, {Fremling}, {Yan}, {Ofek}, {Fransson}, {Filippenko}, {Zheng}, {Brink}, {Copperwheat}, {Foley}, {Brown}, {Siebert}, {Leloudas}, {Cabrera-Lavers}, {Garcia-Alvarez}, {Marante-Barreto}, {Frederick}, {Hung}, {Wheeler}, {Vink{\'o}}, {Thomas}, {Graham}, {Duev}, {Drake}, {Dekany}, {Bellm}, {Rusholme}, {Shupe}, {Andreoni}, {Sharma}, {Riddle}, {van Roestel}, \& {Knezevic}}]{Gal-Yam+2022Natur_AT2019hgp_Icn_SN}
{Gal-Yam}, A., {Bruch}, R., {Schulze}, S., {et~al.} 2022, \nat, 601, 201, \dodoi{10.1038/s41586-021-04155-1}

\bibitem[{{Ginzburg} \& {Balberg}(2014)}]{GinzburgBalberg14}
{Ginzburg}, S., \& {Balberg}, S. 2014, \apj, 780, 18, \dodoi{10.1088/0004-637X/780/1/18}

\bibitem[{{Grefenstette} {et~al.}(2023){Grefenstette}, {Brightman}, {Earnshaw}, {Harrison}, \& {Margutti}}]{2023Grefenstette-ixf}
{Grefenstette}, B.~W., {Brightman}, M., {Earnshaw}, H.~P., {Harrison}, F.~A., \& {Margutti}, R. 2023, \apjl, 952, L3, \dodoi{10.3847/2041-8213/acdf4e}

\bibitem[{{Haynie} \& {Piro}(2021)}]{HayniePiro21CSMBO}
{Haynie}, A., \& {Piro}, A.~L. 2021, \apj, 910, 128, \dodoi{10.3847/1538-4357/abe938}

\bibitem[{{Hosseinzadeh} {et~al.}(2017){Hosseinzadeh}, {Arcavi}, {Valenti}, {McCully}, {Howell}, {Johansson}, {Sollerman}, {Pastorello}, {Benetti}, {Cao}, {Cenko}, {Clubb}, {Corsi}, {Duggan}, {Elias-Rosa}, {Filippenko}, {Fox}, {Fremling}, {Horesh}, {Karamehmetoglu}, {Kasliwal}, {Marion}, {Ofek}, {Sand}, {Taddia}, {Zheng}, {Fraser}, {Gal-Yam}, {Inserra}, {Laher}, {Masci}, {Rebbapragada}, {Smartt}, {Smith}, {Sullivan}, {Surace}, \& {Wo{\'z}niak}}]{Hosseinzadeh+2017_SN_Ibn_diversity}
{Hosseinzadeh}, G., {Arcavi}, I., {Valenti}, S., {et~al.} 2017, \apj, 836, 158, \dodoi{10.3847/1538-4357/836/2/158}

\bibitem[{{IceCube Collaboration}(2013)}]{2013IC-detection}
{IceCube Collaboration}. 2013, Science, 342, 1242856, \dodoi{10.1126/science.1242856}

\bibitem[{{IceCube Collaboration} {et~al.}(2022){IceCube Collaboration}, {Abbasi}, {Ackermann}, {Adams}, {Aguilar}, {Ahlers}, {Ahrens}, {Alameddine}, {Alispach}, {Alves}, {Amin}, {Andeen}, {Anderson}, {Anton}, {Arg{\"u}elles}, {Ashida}, {Axani}, {Bai}, {Balagopal}, {Barbano}, {Barwick}, {Bastian}, {Basu}, {Baur}, {Bay}, {Beatty}, {Becker}, {Becker Tjus}, {Bellenghi}, {Benzvi}, {Berley}, {Bernardini}, {Besson}, {Binder}, {Bindig}, {Blaufuss}, {Blot}, {Boddenberg}, {Bontempo}, {Borowka}, {B{\"o}ser}, {Botner}, {B{\"o}ttcher}, {Bourbeau}, {Bradascio}, {Braun}, {Brinson}, {Bron}, {Brostean-Kaiser}, {Browne}, {Burgman}, {Burley}, {Busse}, {Campana}, {Carnie-Bronca}, {Chen}, {Chen}, {Chirkin}, {Choi}, {Clark}, {Clark}, {Classen}, {Coleman}, {Collin}, {Conrad}, {Coppin}, {Correa}, {Cowen}, {Cross}, {Dappen}, {Dave}, {de Clercq}, {Delaunay}, {Delgado L{\'o}pez}, {Dembinski}, {Deoskar}, {Desai}, {Desiati}, {de Vries}, {de Wasseige}, {de With}, {Deyoung}, {Diaz}, {D{\'\i}az-V{\'e}lez}, {Dittmer}, {Dujmovic}, {Dunkman},
  {Duvernois}, {Dvorak}, {Ehrhardt}, {Eller}, {Engel}, {Erpenbeck}, {Evans}, {Evenson}, {Fan}, {Fazely}, {Fedynitch}, {Feigl}, {Fiedlschuster}, {Fienberg}, {Filimonov}, {Finley}, {Fischer}, {Fox}, {Franckowiak}, {Friedman}, {Fritz}, {F{\"u}rst}, {Gaisser}, {Gallagher}, {Ganster}, {Garcia}, {Garrappa}, {Gerhardt}, {Ghadimi}, {Glaser}, {Glauch}, {Gl{\"u}senkamp}, {Goldschmidt}, {Gonzalez}, {Goswami}, {Grant}, {Gr{\'e}goire}, {Griswold}, {G{\"u}nther}, {Gutjahr}, {Haack}, {Hallgren}, {Halliday}, {Halve}, {Halzen}, {Hanson}, {Hardin}, {Harnisch}, {Haungs}, {Hebecker}, {Helbing}, {Henningsen}, {Hettinger}, {Hickford}, {Hignight}, {Hill}, {Hill}, {Hoffman}, {Hoffmann}, {Hokanson-Fasig}, {Hoshina}, {Huang}, {Huber}, {Huber}, {Hultqvist}, {H{\"u}nnefeld}, {Hussain}, {Hymon}, {in}, {Iovine}, {Ishihara}, {Jansson}, {Japaridze}, {Jeong}, {Jin}, {Jones}, {Kang}, {Kang}, {Kang}, {Kappes}, {Kappesser}, {Kardum}, {Karg}, {Karl}, {Karle}, {Katz}, {Kauer}, {Kellermann}, {Kelley}, {Kheirandish}, {Kin}, {Kintscher}, {Kiryluk},
  {Klein}, {Koirala}, {Kolanoski}, {Kontrimas}, {K{\"o}pke}, {Kopper}, {Kopper}, {Koskinen}, {Koundal}, {Kovacevich}, {Kowalski}, {Kozynets}, {Kun}, {Kurahashi}, {Lad}, {Lagunas Gualda}, {Lanfranchi}, {Larson}, {Lauber}, {Lazar}, {Lee}, {Leonard}, {Leszczy{\'n}ska}, {Li}, {Lincetto}, {Liu}, {Liubarska}, {Lohfink}, {Lozano Mariscal}, {Lu}, {Lucarelli}, {Ludwig}, {Luszczak}, {Lyu}, {Ma}, {Madsen}, {Mahn}, {Makino}, {Mancina}, {Mari{\c{s}}}, {Martinez-Soler}, {Maruyama}, {Mase}, {McElroy}, {McNally}, {Mead}, {Meagher}, {Mechbal}, {Medina}, {Meier}, {Meighen-Berger}, {Micallef}, {Mockler}, {Montaruli}, {Moore}, {Morse}, {Moulai}, {Naab}, {Nagai}, {Nahnhauer}, {Naumann}, {Necker}, {Nguyen}, {Niederhausen}, {Nisa}, {Nowicki}, {Nygren}, {Obertack}, {Pollmann}, {Oehler}, {Oeyen}, {Olivas}, {O'Sullivan}, {Pandya}, {Pankova}, {Park}, {Parker}, {Paudel}, {Paul}, {P{\'e}rez de Los Heros}, {Peters}, {Peterson}, {Philippen}, {Pieper}, {Pittermann}, {Pizzuto}, {Plum}, {Popovych}, {Porcelli}, {Prado Rodriguez}, {Price},
  {Pries}, {Przybylski}, {Rack-Helleis}, {Raissi}, {Rameez}, {Rawlins}, {Rea}, {Rehman}, {Reichherzer}, {Reimann}, {Renzi}, {Resconi}, {Reusch}, {Rhode}, {Richman}, {Riedel}, {Roberts}, {Robertson}, {Roellinghoff}, {Rongen}, {Rott}, {Ruhe}, {Ryckbosch}, {Rysewyk Cantu}, {Safa}, {Saffer}, {Sanchez Herrera}, {Sandrock}, {Sandroos}, {Santander}, {Sarkar}, {Sarkar}, {Satalecka}, {Schaufel}, {Schieler}, {Schindler}, {Schmidt}, {Schneider}, {Schneider}, {Schr{\"o}der}, {Schumacher}, {Schwefer}, {Sclafani}, {Seckel}, {Seunarine}, {Sharma}, {Shefali}, {Silva}, {Skrzypek}, {Smithers}, {Snihur}, {Soedingrekso}, {Soldin}, {Spannfellner}, {Spiczak}, {Spiering}, {Stachurska}, {Stamatikos}, {Stanev}, {Stein}, {Stettner}, {Steuer}, {Stezelberger}, {Stokstad}, {St{\"u}rwald}, {Stuttard}, {Sullivan}, {Taboada}, {Ter-Antonyan}, {Tilav}, {Tischbein}, {Tollefson}, {T{\"o}nnis}, {Toscano}, {Tosi}, {Trettin}, {Tselengidou}, {Tung}, {Turcati}, {Turcotte}, {Turley}, {Twagirayezu}, {Ty}, {Unland Elorrieta}, {Valtonen-Mattila},
  {Vandenbroucke}, {van Eijndhoven}, {Vannerom}, {van Santen}, {Verpoest}, {Walck}, {Watson}, {Weaver}, {Weigel}, {Weindl}, {Weiss}, {Weldert}, {Wendt}, {Werthebach}, {Weyrauch}, {Whitehorn}, {Wiebusch}, {Williams}, {Wolf}, {Woschnagg}, {Wrede}, {Wulff}, {Xu}, {Yanez}, {Yoshida}, {Yu}, {Yuan}, {Zhangan}, \& {Zhelnin}}]{2022IC-1068}
{IceCube Collaboration}, {Abbasi}, R., {Ackermann}, M., {et~al.} 2022, Science, 378, 538, \dodoi{10.1126/science.abg3395}

\bibitem[{{Irani} {et~al.}(2024{\natexlab{a}}){Irani}, {Morag}, {Gal-Yam}, {Waxman}, {Schulze}, {Sollerman}, {Hinds}, {Perley}, {Chen}, {Strotjohann}, {Yaron}, {Zimmerman}, {Bruch}, {Ofek}, {Soumagnac}, {Yang}, {Groom}, {Masci}, {Aubert}, {Riddle}, {Bellm}, \& {Hale}}]{2024Irani-Early-UV-SNII}
{Irani}, I., {Morag}, J., {Gal-Yam}, A., {et~al.} 2024{\natexlab{a}}, \apj, 970, 96, \dodoi{10.3847/1538-4357/ad3de8}

\bibitem[{{Irani} {et~al.}(2024{\natexlab{b}}){Irani}, {Chen}, {Morag}, {Schulze}, {Gal-Yam}, {Strotjohann}, {Yaron}, {Zimmerman}, {Sharon}, {Perley}, {Sollerman}, {Tohuvavohu}, {Das}, {Kasliwal}, {Bruch}, {Brink}, {Zheng}, {Filippenko}, {Patra}, {Vasylyev}, {Yang}, {Graham}, {Bloom}, {Mazzali}, {Purdum}, {Laher}, {Wold}, {Sharma}, {Lacroix}, \& {Medford}}]{irani2024_oqm}
{Irani}, I., {Chen}, P., {Morag}, J., {et~al.} 2024{\natexlab{b}}, \apj, 962, 109, \dodoi{10.3847/1538-4357/ad04d7}

\bibitem[{{Jacobson-Gal{\'a}n} {et~al.}(2022){Jacobson-Gal{\'a}n}, {Dessart}, {Jones}, {Margutti}, {Coppejans}, {Dimitriadis}, {Foley}, {Kilpatrick}, {Matthews}, {Rest}, {Terreran}, {Aleo}, {Auchettl}, {Blanchard}, {Coulter}, {Davis}, {de Boer}, {DeMarchi}, {Drout}, {Earl}, {Gagliano}, {Gall}, {Hjorth}, {Huber}, {Ibik}, {Milisavljevic}, {Pan}, {Rest}, {Ridden-Harper}, {Rojas-Bravo}, {Siebert}, {Smith}, {Taggart}, {Tinyanont}, {Wang}, \& {Zenati}}]{Jacobson-Galan+2022ApJ_SN2020tlf_precursor_MassLoss}
{Jacobson-Gal{\'a}n}, W.~V., {Dessart}, L., {Jones}, D.~O., {et~al.} 2022, \apj, 924, 15, \dodoi{10.3847/1538-4357/ac3f3a}

\bibitem[{{Jacobson-Gal{\'a}n} {et~al.}(2023){Jacobson-Gal{\'a}n}, {Dessart}, {Margutti}, {Chornock}, {Foley}, {Kilpatrick}, {Jones}, {Taggart}, {Angus}, {Bhattacharjee}, {Braff}, {Brethauer}, {Burgasser}, {Cao}, {Carlile}, {Chambers}, {Coulter}, {Dominguez-Ruiz}, {Dickinson}, {de Boer}, {Gagliano}, {Gall}, {Gao}, {Gates}, {Gomez}, {Guolo}, {Halford}, {Hjorth}, {Huber}, {Johnson}, {Karpoor}, {Laskar}, {LeBaron}, {Li}, {Lin}, {Loch}, {Lynam}, {Magnier}, {Maloney}, {Matthews}, {McDonald}, {Miao}, {Milisavljevic}, {Pan}, {Pradyumna}, {Ransome}, {Rees}, {Rest}, {Rojas-Bravo}, {Sandford}, {Ascencio}, {Sanjaripour}, {Savino}, {Sears}, {Sharei}, {Smartt}, {Softich}, {Theissen}, {Tinyanont}, {Tohfa}, {Villar}, {Wang}, {Wainscoat}, {Westerling}, {Wiston}, {Wozniak}, {Yadavalli}, \& {Zenati}}]{2023Jacobson-ixf}
{Jacobson-Gal{\'a}n}, W.~V., {Dessart}, L., {Margutti}, R., {et~al.} 2023, \apjl, 954, L42, \dodoi{10.3847/2041-8213/acf2ec}

\bibitem[{{Jacobson-Gal{\'a}n} {et~al.}(2024){Jacobson-Gal{\'a}n}, {Dessart}, {Davis}, {Kilpatrick}, {Margutti}, {Foley}, {Chornock}, {Terreran}, {Hiramatsu}, {Newsome}, {Padilla Gonzalez}, {Pellegrino}, {Howell}, {Filippenko}, {Anderson}, {Angus}, {Auchettl}, {Bostroem}, {Brink}, {Cartier}, {Coulter}, {de Boer}, {Drout}, {Earl}, {Ertini}, {Farah}, {Farias}, {Gall}, {Gao}, {Gerlach}, {Guo}, {Haynie}, {Hosseinzadeh}, {Ibik}, {Jha}, {Jones}, {Langeroodi}, {LeBaron}, {Magnier}, {Piro}, {Raimundo}, {Rest}, {Rest}, {Rich}, {Rojas-Bravo}, {Sears}, {Taggart}, {Villar}, {Wainscoat}, {Wang}, {Wasserman}, {Yan}, {Yang}, {Zhang}, \& {Zheng}}]{Jacobson-Galan2024}
{Jacobson-Gal{\'a}n}, W.~V., {Dessart}, L., {Davis}, K.~W., {et~al.} 2024, \apj, 970, 189, \dodoi{10.3847/1538-4357/ad4a2a}

\bibitem[{{Kalashev} {et~al.}(2015){Kalashev}, {Semikoz}, \& {Tkachev}}]{2015Semikoz-AGN-core}
{Kalashev}, O., {Semikoz}, D., \& {Tkachev}, I. 2015, Soviet Journal of Experimental and Theoretical Physics, 120, 541, \dodoi{10.1134/S106377611503022X}

\bibitem[{{Karamehmetoglu} {et~al.}(2021){Karamehmetoglu}, {Fransson}, {Sollerman}, {Tartaglia}, {Taddia}, {De}, {Fremling}, {Bagdasaryan}, {Barbarino}, {Bellm}, {Dekany}, {Dugas}, {Giomi}, {Goobar}, {Graham}, {Ho}, {Laher}, {Masci}, {Neill}, {Perley}, {Riddle}, {Rusholme}, \& {Soumagnac}}]{Karamehmetoglu21}
{Karamehmetoglu}, E., {Fransson}, C., {Sollerman}, J., {et~al.} 2021, \aap, 649, A163, \dodoi{10.1051/0004-6361/201936308}

\bibitem[{{Katz} {et~al.}(2011){Katz}, {Sapir}, \& {Waxman}}]{2011KSW11}
{Katz}, B., {Sapir}, N., \& {Waxman}, E. 2011, arXiv e-prints, arXiv:1106.1898, \dodoi{10.48550/arXiv.1106.1898}

\bibitem[{{Khazov} {et~al.}(2016){Khazov}, {Yaron}, {Gal-Yam}, {Manulis}, {Rubin}, {Kulkarni}, {Arcavi}, {Kasliwal}, {Ofek}, {Cao}, {Perley}, {Sollerman}, {Horesh}, {Sullivan}, {Filippenko}, {Nugent}, {Howell}, {Cenko}, {Silverman}, {Ebeling}, {Taddia}, {Johansson}, {Laher}, {Surace}, {Rebbapragada}, {Wozniak}, \& {Matheson}}]{Khazov+2016_FlashSearchSample}
{Khazov}, D., {Yaron}, O., {Gal-Yam}, A., {et~al.} 2016, \apj, 818, 3, \dodoi{10.3847/0004-637X/818/1/3}

\bibitem[{{Kuriyama} \& {Shigeyama}(2020)}]{2020Kuriyama-CSM-profile}
{Kuriyama}, N., \& {Shigeyama}, T. 2020, \aap, 635, A127, \dodoi{10.1051/0004-6361/201937226}

\bibitem[{{Levinson} \& {Nakar}(2020)}]{2020Levinson-Nakar-rev}
{Levinson}, A., \& {Nakar}, E. 2020, \physrep, 866, 1, \dodoi{10.1016/j.physrep.2020.04.003}

\bibitem[{{Li} {et~al.}(2011){Li}, {Leaman}, {Chornock}, {Filippenko}, {Poznanski}, {Ganeshalingam}, {Wang}, {Modjaz}, {Jha}, {Foley}, \& {Smith}}]{Li+2011MNRAS_NearbySN_Rates_ObservedLuminomsityFunction}
{Li}, W., {Leaman}, J., {Chornock}, R., {et~al.} 2011, \mnras, 412, 1441, \dodoi{10.1111/j.1365-2966.2011.18160.x}

\bibitem[{{Li}(2019)}]{2019Zhuo-PeV-BO}
{Li}, Z. 2019, Science China Physics, Mechanics, and Astronomy, 62, 959511, \dodoi{10.1007/s11433-018-9350-3}

\bibitem[{{Margalit}(2022)}]{MargalitThickCSM22}
{Margalit}, B. 2022, \apj, 933, 238, \dodoi{10.3847/1538-4357/ac771a}

\bibitem[{{Marshall} {et~al.}(2004){Marshall}, {van Loon}, {Matsuura}, {Wood}, {Zijlstra}, \& {Whitelock}}]{2004Marshall-Mdot}
{Marshall}, J.~R., {van Loon}, J.~T., {Matsuura}, M., {et~al.} 2004, \mnras, 355, 1348, \dodoi{10.1111/j.1365-2966.2004.08417.x}

\bibitem[{{Mart{\'\i}-Devesa} {et~al.}(2024){Mart{\'\i}-Devesa}, {Cheung}, {Di Lalla}, {Renaud}, {Principe}, {Omodei}, \& {Acero}}]{Marti-2024A&A_SN2023ixf_FermiObservations}
{Mart{\'\i}-Devesa}, G., {Cheung}, C.~C., {Di Lalla}, N., {et~al.} 2024, \aap, 686, A254, \dodoi{10.1051/0004-6361/202349061}

\bibitem[{{Matzner} \& {McKee}(1999)}]{1999MnM-SN-shock}
{Matzner}, C.~D., \& {McKee}, C.~F. 1999, \apj, 510, 379, \dodoi{10.1086/306571}

\bibitem[{{M{\'e}sz{\'a}ros} \& {Waxman}(2001)}]{2001MW-chocked}
{M{\'e}sz{\'a}ros}, P., \& {Waxman}, E. 2001, \prl, 87, 171102, \dodoi{10.1103/PhysRevLett.87.171102}

\bibitem[{{Morag} {et~al.}(2023){Morag}, {Sapir}, \& {Waxman}}]{2023Morag-RSG-I}
{Morag}, J., {Sapir}, N., \& {Waxman}, E. 2023, \mnras, 522, 2764, \dodoi{10.1093/mnras/stad899}

\bibitem[{{Moriya} {et~al.}(2013){Moriya}, {Blinnikov}, {Baklanov}, {Sorokina}, \& {Dolgov}}]{Moriya13}
{Moriya}, T.~J., {Blinnikov}, S.~I., {Baklanov}, P.~V., {Sorokina}, E.~I., \& {Dolgov}, A.~D. 2013, \mnras, 430, 1402, \dodoi{10.1093/mnras/stt011}

\bibitem[{{Morozova} {et~al.}(2018){Morozova}, {Piro}, \& {Valenti}}]{2018Morozova-SNII-Mloss-lightcurve}
{Morozova}, V., {Piro}, A.~L., \& {Valenti}, S. 2018, \apj, 858, 15, \dodoi{10.3847/1538-4357/aab9a6}

\bibitem[{{Murase}(2024)}]{2024Murase-SNe}
{Murase}, K. 2024, \prd, 109, 103020, \dodoi{10.1103/PhysRevD.109.103020}

\bibitem[{{Murase} {et~al.}(2016){Murase}, {Guetta}, \& {Ahlers}}]{2016Murase-g-tension}
{Murase}, K., {Guetta}, D., \& {Ahlers}, M. 2016, \prl, 116, 071101, \dodoi{10.1103/PhysRevLett.116.071101}

\bibitem[{{Murase} \& {Ioka}(2013)}]{2013Murase-chocked}
{Murase}, K., \& {Ioka}, K. 2013, \prl, 111, 121102, \dodoi{10.1103/PhysRevLett.111.121102}

\bibitem[{{Murase} {et~al.}(2020){Murase}, {Kimura}, \& {M{\'e}sz{\'a}ros}}]{2020Murase-AGN-core}
{Murase}, K., {Kimura}, S.~S., \& {M{\'e}sz{\'a}ros}, P. 2020, \prl, 125, 011101, \dodoi{10.1103/PhysRevLett.125.011101}

\bibitem[{{Murase} {et~al.}(2011){Murase}, {Thompson}, {Lacki}, \& {Beacom}}]{2011Murase-SNe}
{Murase}, K., {Thompson}, T.~A., {Lacki}, B.~C., \& {Beacom}, J.~F. 2011, \prd, 84, 043003, \dodoi{10.1103/PhysRevD.84.043003}

\bibitem[{{Ofek} {et~al.}(2023){Ofek}, {Ben-Ami}, {Polishook}, {Segre}, {Blumenzweig}, {et~al.}}]{Ofek+2023PASP_LAST_Overview}
{Ofek}, E.~O., {Ben-Ami}, S., {Polishook}, D., {et~al.} 2023, \pasp, 135, 065001, \dodoi{10.1088/1538-3873/acd8f0}

\bibitem[{{Ofek} {et~al.}(2010){Ofek}, {Rabinak}, {Neill}, {Arcavi}, {Cenko}, {Waxman}, {Kulkarni}, {Gal-Yam}, {Nugent}, {Bildsten}, {Bloom}, {Filippenko}, {Forster}, {Howell}, {Jacobsen}, {Kasliwal}, {Law}, {Martin}, {Poznanski}, {Quimby}, {Shen}, {Sullivan}, {Dekany}, {Rahmer}, {Hale}, {Smith}, {Zolkower}, {Velur}, {Walters}, {Henning}, {Bui}, \& {McKenna}}]{2010Ofek-wind-BO}
{Ofek}, E.~O., {Rabinak}, I., {Neill}, J.~D., {et~al.} 2010, \apj, 724, 1396, \dodoi{10.1088/0004-637X/724/2/1396}

\bibitem[{{Ofek} {et~al.}(2014{\natexlab{a}}){Ofek}, {Sullivan}, {Shaviv}, {Steinbok}, {Arcavi}, {Gal-Yam}, {Tal}, {Kulkarni}, {Nugent}, \& {Ben-Ami}}]{Ofek+2014_IIn_precursors}
{Ofek}, E.~O., {Sullivan}, M., {Shaviv}, N.~J., {et~al.} 2014{\natexlab{a}}, \apj, 789, 104, \dodoi{10.1088/0004-637X/789/2/104}

\bibitem[{{Ofek} {et~al.}(2014{\natexlab{b}}){Ofek}, {Zoglauer}, {Boggs}, {Barri{\'e}re}, {Reynolds}, {Fryer}, {Harrison}, {Cenko}, {Kulkarni}, \& {Gal-Yam}}]{Ofek+2014_NuSTAR}
{Ofek}, E.~O., {Zoglauer}, A., {Boggs}, S.~E., {et~al.} 2014{\natexlab{b}}, \apj, 781, 42, \dodoi{10.1088/0004-637X/781/1/42}

\bibitem[{{Padovani} {et~al.}(2024){Padovani}, {Resconi}, {Ajello}, {Bellenghi}, {Bianchi}, {Blasi}, {Huang}, {Gabici}, {G{\'a}mez Rosas}, {Niederhausen}, {Peretti}, {Eichmann}, {Guetta}, {Lamastra}, \& {Shimizu}}]{2024Resconi-1068}
{Padovani}, P., {Resconi}, E., {Ajello}, M., {et~al.} 2024, arXiv e-prints, arXiv:2405.20146, \dodoi{10.48550/arXiv.2405.20146}

\bibitem[{{Pastorello} {et~al.}(2007){Pastorello}, {Smartt}, {Mattila}, {Eldridge}, {Young}, {Itagaki}, {Yamaoka}, {Navasardyan}, {Valenti}, {Patat}, {Agnoletto}, {Augusteijn}, {Benetti}, {Cappellaro}, {Boles}, {Bonnet-Bidaud}, {Botticella}, {Bufano}, {Cao}, {Deng}, {Dennefeld}, {Elias-Rosa}, {Harutyunyan}, {Keenan}, {Iijima}, {Lorenzi}, {Mazzali}, {Meng}, {Nakano}, {Nielsen}, {Smoker}, {Stanishev}, {Turatto}, {Xu}, \& {Zampieri}}]{Pastorello+2007_SN2006jc_precursor}
{Pastorello}, A., {Smartt}, S.~J., {Mattila}, S., {et~al.} 2007, \nat, 447, 829, \dodoi{10.1038/nature05825}

\bibitem[{{Pastorello} {et~al.}(2008){Pastorello}, {Mattila}, {Zampieri}, {Della Valle}, {Smartt}, {Valenti}, {Agnoletto}, {Benetti}, {Benn}, {Branch}, {Cappellaro}, {Dennefeld}, {Eldridge}, {Gal-Yam}, {Harutyunyan}, {Hunter}, {Kjeldsen}, {Lipkin}, {Mazzali}, {Milne}, {Navasardyan}, {Ofek}, {Pian}, {Shemmer}, {Spiro}, {Stathakis}, {Taubenberger}, {Turatto}, \& {Yamaoka}}]{Pastorello08}
{Pastorello}, A., {Mattila}, S., {Zampieri}, L., {et~al.} 2008, \mnras, 389, 113, \dodoi{10.1111/j.1365-2966.2008.13602.x}

\bibitem[{{Pellegrino} {et~al.}(2022){Pellegrino}, {Howell}, {Terreran}, {Arcavi}, {Bostroem}, {Brown}, {Burke}, {Dong}, {Gilkis}, {Hiramatsu}, {Hosseinzadeh}, {McCully}, {Modjaz}, {Newsome}, {Gonzalez}, {Pritchard}, {Sand}, {Valenti}, \& {Williamson}}]{Pellegrino22}
{Pellegrino}, C., {Howell}, D.~A., {Terreran}, G., {et~al.} 2022, \apj, 938, 73, \dodoi{10.3847/1538-4357/ac8ff6}

\bibitem[{{Perley} {et~al.}(2022){Perley}, {Sollerman}, {Schulze}, {Yao}, {Fremling}, {Gal-Yam}, {Ho}, {Yang}, {Kool}, {Irani}, {Yan}, {Andreoni}, {Baade}, {Bellm}, {Brink}, {Chen}, {Cikota}, {Coughlin}, {Dahiwale}, {Dekany}, {Duev}, {Filippenko}, {Hoeflich}, {Kasliwal}, {Kulkarni}, {Lunnan}, {Masci}, {Maund}, {Medford}, {Riddle}, {Rosnet}, {Shupe}, {Strotjohann}, {Tzanidakis}, \& {Zheng}}]{Perley22}
{Perley}, D.~A., {Sollerman}, J., {Schulze}, S., {et~al.} 2022, \apj, 927, 180, \dodoi{10.3847/1538-4357/ac478e}

\bibitem[{{Petropoulou} {et~al.}(2017){Petropoulou}, {Coenders}, {Vasilopoulos}, {Kamble}, \& {Sironi}}]{2017Petropoulou-IIn-Nu}
{Petropoulou}, M., {Coenders}, S., {Vasilopoulos}, G., {Kamble}, A., \& {Sironi}, L. 2017, \mnras, 470, 1881, \dodoi{10.1093/mnras/stx1251}

\bibitem[{{Roth} \& {Kasen}(2015)}]{KasenNumeric15}
{Roth}, N., \& {Kasen}, D. 2015, \apjs, 217, 9, \dodoi{10.1088/0067-0049/217/1/9}

\bibitem[{{Sarmah} {et~al.}(2022){Sarmah}, {Chakraborty}, {Tamborra}, \& {Auchettl}}]{2022Tamborra-SNe}
{Sarmah}, P., {Chakraborty}, S., {Tamborra}, I., \& {Auchettl}, K. 2022, \jcap, 2022, 011, \dodoi{10.1088/1475-7516/2022/08/011}

\bibitem[{{Schlegel}(1990)}]{Schlegel1990}
{Schlegel}, E.~M. 1990, \mnras, 244, 269

\bibitem[{{Shiode} \& {Quataert}(2014)}]{Shiode+Quataert2014_WaveDrivenMassLoss}
{Shiode}, J.~H., \& {Quataert}, E. 2014, \apj, 780, 96, \dodoi{10.1088/0004-637X/780/1/96}

\bibitem[{{Shvartzvald} {et~al.}(2024){Shvartzvald}, {Waxman}, {Gal-Yam}, {Ofek}, {Ben-Ami}, {Berge}, {Kowalski}, {B{\"u}hler}, {Worm}, {Rhoads}, {Arcavi}, {Maoz}, {Polishook}, {Stone}, {Trakhtenbrot}, {Ackermann}, {Aharonson}, {Birnholtz}, {Chelouche}, {Guetta}, {Hallakoun}, {Horesh}, {Kushnir}, {Mazeh}, {Nordin}, {Ofir}, {Ohm}, {Parsons}, {Pe'er}, {Perets}, {Perdelwitz}, {Poznanski}, {Sadeh}, {Sagiv}, {Shahaf}, {Soumagnac}, {Tal-Or}, {Santen}, {Zackay}, {Guttman}, {Rekhi}, {Townsend}, {Weinstein}, \& {Wold}}]{2024Shvartzvald-ULTRASAT}
{Shvartzvald}, Y., {Waxman}, E., {Gal-Yam}, A., {et~al.} 2024, \apj, 964, 74, \dodoi{10.3847/1538-4357/ad2704}

\bibitem[{{Smith}(2014)}]{Smith2014ARA&A_TypeIIn_SN_MassLoss}
{Smith}, N. 2014, \araa, 52, 487, \dodoi{10.1146/annurev-astro-081913-040025}

\bibitem[{{Smith} \& {Arnett}(2014)}]{2014Smith-Mloss}
{Smith}, N., \& {Arnett}, W.~D. 2014, \apj, 785, 82, \dodoi{10.1088/0004-637X/785/2/82}

\bibitem[{{Soker} \& {Kashi}(2013)}]{Soker13CommonE}
{Soker}, N., \& {Kashi}, A. 2013, \apjl, 764, L6, \dodoi{10.1088/2041-8205/764/1/L6}

\bibitem[{{Sridhar} {et~al.}(2024){Sridhar}, {Metzger}, \& {Fang}}]{2024Metzger-FRB-nu}
{Sridhar}, N., {Metzger}, B.~D., \& {Fang}, K. 2024, \apj, 960, 74, \dodoi{10.3847/1538-4357/ad03e8}

\bibitem[{{Stecker} {et~al.}(1991){Stecker}, {Done}, {Salamon}, \& {Sommers}}]{1991Stecker-AGN-core}
{Stecker}, F.~W., {Done}, C., {Salamon}, M.~H., \& {Sommers}, P. 1991, \prl, 66, 2697, \dodoi{10.1103/PhysRevLett.66.2697}

\bibitem[{{Steeghs} {et~al.}(2022){Steeghs}, {Galloway}, {Ackley}, {Dyer}, {Lyman}, {Ulaczyk}, {Cutter}, {Mong}, {Dhillon}, {O'Brien}, {Ramsay}, {Poshyachinda}, {Kotak}, {Nuttall}, {Pall{\'e}}, {Breton}, {Pollacco}, {Thrane}, {Aukkaravittayapun}, {Awiphan}, {Burhanudin}, {Chote}, {Chrimes}, {Daw}, {Duffy}, {Eyles-Ferris}, {Gompertz}, {Heikkil{\"a}}, {Irawati}, {Kennedy}, {Killestein}, {Kuncarayakti}, {Levan}, {Littlefair}, {Makrygianni}, {Marsh}, {Mata-Sanchez}, {Mattila}, {Maund}, {McCormac}, {Mkrtichian}, {Mullaney}, {Noysena}, {Patel}, {Rol}, {Sawangwit}, {Stanway}, {Starling}, {Str{\o}m}, {Tooke}, {West}, {White}, \& {Wiersema}}]{GOTO2022}
{Steeghs}, D., {Galloway}, D.~K., {Ackley}, K., {et~al.} 2022, \mnras, 511, 2405, \dodoi{10.1093/mnras/stac013}

\bibitem[{{Strotjohann} {et~al.}(2021){Strotjohann}, {Ofek}, {Gal-Yam}, {Bruch}, {Schulze}, {Shaviv}, {Sollerman}, {Filippenko}, {Yaron}, {Fremling}, {Nordin}, {Kool}, {Perley}, {Ho}, {Yang}, {Yao}, {Soumagnac}, {Graham}, {Barbarino}, {Tartaglia}, {De}, {Goldstein}, {Cook}, {Brink}, {Taggart}, {Yan}, {Lunnan}, {Kasliwal}, {Kulkarni}, {Nugent}, {Masci}, {Rosnet}, {Adams}, {Andreoni}, {Bagdasaryan}, {Bellm}, {Burdge}, {Duev}, {Dugas}, {Frederick}, {Goldwasser}, {Hankins}, {Irani}, {Karambelkar}, {Kupfer}, {Liang}, {Neill}, {Porter}, {Riddle}, {Sharma}, {Short}, {Taddia}, {Tzanidakis}, {van Roestel}, {Walters}, \& {Zhuang}}]{2021Nora-IIn-precursors}
{Strotjohann}, N.~L., {Ofek}, E.~O., {Gal-Yam}, A., {et~al.} 2021, \apj, 907, 99, \dodoi{10.3847/1538-4357/abd032}

\bibitem[{{Su{\'a}rez-Madrigal} {et~al.}(2013){Su{\'a}rez-Madrigal}, {Krumholz}, \& {Ramirez-Ruiz}}]{Suarez13Mdot}
{Su{\'a}rez-Madrigal}, A., {Krumholz}, M., \& {Ramirez-Ruiz}, E. 2013, arXiv e-prints, arXiv:1304.2317.
\newblock \doarXiv{1304.2317}

\bibitem[{{Svirski} \& {Nakar}(2014)}]{SvirskiNakar14}
{Svirski}, G., \& {Nakar}, E. 2014, \apj, 788, 113, \dodoi{10.1088/0004-637X/788/2/113}

\bibitem[{{Terreran} {et~al.}(2022){Terreran}, {Jacobson-Gal{\'a}n}, {Groh}, {Margutti}, {Coppejans}, {Dimitriadis}, {Kilpatrick}, {Matthews}, {Siebert}, {Angus}, {Brink}, {Filippenko}, {Foley}, {Jones}, {Tinyanont}, {Gall}, {Pfister}, {Zenati}, {Ansari}, {Auchettl}, {El-Badry}, {Magnier}, \& {Zheng}}]{Terreran2022}
{Terreran}, G., {Jacobson-Gal{\'a}n}, W.~V., {Groh}, J.~H., {et~al.} 2022, \apj, 926, 20, \dodoi{10.3847/1538-4357/ac3820}

\bibitem[{{Thwaites} {et~al.}(2023){Thwaites}, {Vandenbroucke}, {Santander}, \& {IceCube Collaboration}}]{2023ATel16043_SN2023ixf_IceCubeLimits}
{Thwaites}, J., {Vandenbroucke}, J., {Santander}, M., \& {IceCube Collaboration}. 2023, The Astronomer's Telegram, 16043, 1

\bibitem[{{Tsang} {et~al.}(2022){Tsang}, {Kasen}, \& {Bildsten}}]{2022Kasen-CSM-profile}
{Tsang}, B. T.~H., {Kasen}, D., \& {Bildsten}, L. 2022, \apj, 936, 28, \dodoi{10.3847/1538-4357/ac83bc}

\bibitem[{{Tsuna} {et~al.}(2021){Tsuna}, {Takei}, {Kuriyama}, \& {Shigeyama}}]{2021Tsuna-CSM-profile}
{Tsuna}, D., {Takei}, Y., {Kuriyama}, N., \& {Shigeyama}, T. 2021, \pasj, 73, 1128, \dodoi{10.1093/pasj/psab063}

\bibitem[{{van Loon} {et~al.}(2005){van Loon}, {Cioni}, {Zijlstra}, \& {Loup}}]{2005Loon-Mdot}
{van Loon}, J.~T., {Cioni}, M. R.~L., {Zijlstra}, A.~A., \& {Loup}, C. 2005, \aap, 438, 273, \dodoi{10.1051/0004-6361:20042555}

\bibitem[{{Waxman}(2013)}]{2013WB}
{Waxman}, E. 2013, arXiv e-prints, arXiv:1312.0558, \dodoi{10.48550/arXiv.1312.0558}

\bibitem[{{Waxman} \& {Bahcall}(1998)}]{1998WB}
{Waxman}, E., \& {Bahcall}, J. 1998, \prd, 59, 023002, \dodoi{10.1103/PhysRevD.59.023002}

\bibitem[{{Waxman} \& {Katz}(2017)}]{2017WK-SN-book}
{Waxman}, E., \& {Katz}, B. 2017, in Handbook of Supernovae, ed. A.~W. {Alsabti} \& P.~{Murdin}, 967, \dodoi{10.1007/978-3-319-21846-5_33}

\bibitem[{{Weaver}(1976)}]{Weaver1976_StructureOfSupernovaeShocks}
{Weaver}, T.~A. 1976, \apjs, 32, 233, \dodoi{10.1086/190398}

\bibitem[{{Woosley} {et~al.}(2007){Woosley}, {Blinnikov}, \& {Heger}}]{2007Woosley-PI-Mloss}
{Woosley}, S.~E., {Blinnikov}, S., \& {Heger}, A. 2007, \nat, 450, 390, \dodoi{10.1038/nature06333}

\bibitem[{{Yaron} {et~al.}(2017){Yaron}, {Perley}, {Gal-Yam}, {Groh}, {Horesh}, {Ofek}, {Kulkarni}, {Sollerman}, {Fransson}, {Rubin}, {Szabo}, {Sapir}, {Taddia}, {Cenko}, {Valenti}, {Arcavi}, {Howell}, {Kasliwal}, {Vreeswijk}, {Khazov}, {Fox}, {Cao}, {Gnat}, {Kelly}, {Nugent}, {Filippenko}, {Laher}, {Wozniak}, {Lee}, {Rebbapragada}, {Maguire}, {Sullivan}, \& {Soumagnac}}]{Yaron+2017_PTF13dqy_HighIonization_FlashSPectroscopy}
{Yaron}, O., {Perley}, D.~A., {Gal-Yam}, A., {et~al.} 2017, Nature Physics, 13, 510, \dodoi{10.1038/nphys4025}

\bibitem[{{Zegarelli} {et~al.}(2024){Zegarelli}, {Guetta}, {Celli}, {Gagliardini}, {Di Palma}, \& {Bartos}}]{2024Zegarelli-choked-jets}
{Zegarelli}, A., {Guetta}, D., {Celli}, S., {et~al.} 2024, arXiv e-prints, arXiv:2403.16234, \dodoi{10.48550/arXiv.2403.16234}

\bibitem[{{Zhang} {et~al.}(2020){Zhang}, {Wang}, {J{\'o}zsef}, {Zhai}, {Zhang}, {Filippenko}, {Brink}, {Zheng}, {Wyrzykowski}, {Miko{\l}ajczyk}, {Huang}, {Rui}, {Mo}, {Sai}, {Zhang}, {Wang}, {DerKacy}, {Baron}, {S{\'a}rneczky}, {B{\'o}di}, {Cs{\"o}rnyei}, {Hanyecz}, {Ign{\'a}cz}, {Kalup}, {Kriskovics}, {K{\"o}nyves-T{\'o}th}, {Ordasi}, {P{\'a}l}, {S{\'o}dor}, {Szak{\'a}ts}, {Vida}, \& {Zsidi}}]{Zhang2020}
{Zhang}, J., {Wang}, X., {J{\'o}zsef}, V., {et~al.} 2020, \mnras, 498, 84, \dodoi{10.1093/mnras/staa2273}

\bibitem[{{Zimmerman} {et~al.}(2024){Zimmerman}, {Irani}, {Chen}, {Gal-Yam}, {Schulze}, {Perley}, {Sollerman}, {Filippenko}, {Shenar}, {Yaron}, {Shahaf}, {Bruch}, {Ofek}, {De Cia}, {Brink}, {Yang}, {Vasylyev}, {Ben Ami}, {Aubert}, {Badash}, {Bloom}, {Brown}, {De}, {Dimitriadis}, {Fransson}, {Fremling}, {Hinds}, {Horesh}, {Johansson}, {Kasliwal}, {Kulkarni}, {Kushnir}, {Martin}, {Matuzewski}, {McGurk}, {Miller}, {Morag}, {Neil}, {Nugent}, {Post}, {Prusinski}, {Qin}, {Raichoor}, {Riddle}, {Rowe}, {Rusholme}, {Sfaradi}, {Sjoberg}, {Soumagnac}, {Stein}, {Strotjohann}, {Terwel}, {Wasserman}, {Wise}, {Wold}, {Yan}, \& {Zhang}}]{2024Zimmerman-ixf}
{Zimmerman}, E.~A., {Irani}, I., {Chen}, P., {et~al.} 2024, \nat, 627, 759, \dodoi{10.1038/s41586-024-07116-6}

\end{thebibliography}

\end{document}